\documentclass[twocolumn]{aastex631}
\usepackage{soul}
\usepackage{comment}
\usepackage{amsmath}
\usepackage{amssymb}
\usepackage{bbold}
\usepackage[super]{nth}

\newcommand\bb[1]{\mbox{\boldmath{$#1$}}}
\newcommand\bdbldot{\,\bb{:}\,}

\begin{document}

\title{Alpha Core-Beam Origin in Low-$\beta$ Solar Wind Plasma: Insights from Fully Kinetic Simulation}

%\title{Nonlinear Landau Damping as a Possible Formation Mechanism of Alpha-Beam Species in Low-$\beta$ Solar Wind}

\author[0000-0001-5079-7941]{L. Pezzini}
\affiliation{Centre for mathematical Plasma Astrophysics, Department of Mathematics, KU Leuven, Celestijnenlaan 200B, B-3001 Leuven, Belgium}
\affiliation{Solar-Terrestrial Centre of Excellence--SIDC, Royal Observatory of Belgium, 1180 Brussels, Belgium}

\author[0000-0002-7526-8154]{F. Bacchini}
\affiliation{Centre for mathematical Plasma Astrophysics, Department of Mathematics, KU Leuven, Celestijnenlaan 200B, B-3001 Leuven, Belgium}
\affiliation{Solar-Terrestrial Center of Excellence and Space Physics, Royal Belgian Institute for Space Aeronomy, B-1180 Brussels,
Belgium}

\author[0000-0002-2542-9810]{A. N. Zhukov}
\affiliation{Solar-Terrestrial Centre of Excellence--SIDC, Royal Observatory of Belgium, 1180 Brussels, Belgium}
\affiliation{Skobeltsyn Institute of Nuclear Physics, Moscow State University, 119991 Moscow, Russia}

\author[0000-0001-7233-2555]{G. Arrò}
\affiliation{Department of Physics, University of Wisconsin-Madison, Madison, WI 53706, USA}

\author[0000-0003-3223-1498]{R. A. López}
\affiliation{Research Center in the intersection of Plasma Physics, Matter, and Complexity ($P^2 mc$),\\ Comisi\'on Chilena de Energ\'{\i}a Nuclear, Casilla 188-D, Santiago, Chile}
\affiliation{Departamento de Ciencias F\'{\i}sicas, Facultad de Ciencias Exactas, Universidad Andres Bello, Sazi\'e 2212, Santiago 8370136, Chile}
%–––––––––––––––––––––––––––––––––––––––––––––––––––––––––––––––––––%
%                         ABSTRACT SECTION                        %
%–––––––––––––––––––––––––––––––––––––––––––––––––––––––––––––––––––%

\begin{abstract}

In-situ observations of the fast solar wind in the inner-heliosphere show that minor ions and ion sub-populations often exhibit distinct drift velocities. Both alpha particles and proton beams stream at speeds that rarely exceed the local Alfv\'{e}n speed relative to the core protons, suggesting the presence of instabilities that constrain their maximum drift. We aim to propose a mechanism that generates an alpha-particle beam through non-linear Landau damping, primarily driven by the relative super-Alfv\'{e}nic drift between protons and alpha particles. To investigate this process, we perform one-dimensional, fully kinetic particle-in-cell simulations of a non-equilibrium multi-species plasma, complemented by its linear theory to validate the model during the linear phase. Our results provide clear evidence that the system evolves by producing an alpha-particle beam, thereby suggesting a local mechanism for alpha-beam generation via non-linear Landau damping.

\end{abstract}

\keywords{Solar wind (1534)--Space plasmas (1544)--Plasma astrophysics (1261)--Plasma physics (2089)}

%\keywords{\uat{Solar wind}{1534}---\uat{Space plasmas}{1544}---\uat{Plasma physics}{2089}}

%–––––––––––––––––––––––––––––––––––––––––––––––––––––––––––––––––––%
%                        INTRODUCTION SECTION                       %
%–––––––––––––––––––––––––––––––––––––––––––––––––––––––––––––––––––%

\section{Introduction} \label{sec:intro}
The Sun continuously loses mass and energy through an outflow of magnetised plasma known as the solar wind (SW), which is radially expanding into the outer space, defining the heliosphere \citep{marsch2006}. SW is particularly intriguing because it simultaneously accelerates outward from the Sun and exhibits temperatures significantly higher than the solar surface itself \citep{cranmer2019}--a counterintuitive characteristic for an expanding plasma. This discrepancy is commonly referred to as the SW accelerating and heating problem. To address these long-standing questions, the Parker Solar Probe (PSP) \citep{fox2016} and Solar Orbiter (SolO) \citep{muller2020} missions were launched with the goal of providing unprecedented insights into the underlying physical processes.

The SW plasma is primarily composed of electrons, protons (singly ionized hydrogen, $\mathrm{H^{+}}$), and alpha particles (doubly ionized helium, $\mathrm{He^{2+}}$). Alpha particles are the most significant minor ion species, contributing approximately $15$–$20\%$ of the total mass density and thereby exerting a crucial influence on the SW’s dynamics and thermodynamics \citep{bame1977, li2006, marsch1984, pizzo1983}. Measurements of protons and alpha particles in the SW consistently indicate that their temperatures decrease less rapidly with increasing heliocentric distance than predicted by adiabatic or double-adiabatic expansion models, thereby providing evidence for the presence of an additional heating mechanism \citep{chew1956, cranmer2009, gazis1982, hellinger2011, hellinger2013, lamarche2014, marsch1982b, marsch1982,  marsch1983, maruca2011, miyake1987, richardson1995, schwartz1983, thieme1989}.

Plasma is inherently kinetic in nature: on the microscopic scale, it comprises charged particles that interact self-consistently with electromagnetic fields. In particular, SW plasma is weakly collisional, as the typical ion-ion collision timescale is much longer than the travel time from the Sun \citep{kasper2008}. At the same time, the SW is intrinsically multiscale, with processes such as expansion and turbulence redistributing energy across different scales \citep{arro2022, chandran2019, cerri2019, howes2008, matthaeus2020, verscharen2019}. These processes drive non-equilibrium features in the distribution functions of the particle species \citep{goldstein2000, kasper2013, marsch1982b, marsch1982, maruca2012, reisenfeld2001}, resulting in symmetry-breaking characteristics of the velocity distribution functions (VDFs). Such non-equilibrium features include relative drifts between different plasma species and same plasma species (beam) along the direction of $\boldsymbol{B}$ background magnetic field, as well as temperature anisotropies with respect to the same field \citep{afify2025, bonhome2025, klein2021, micera2020, micera2020a, micera2021, ofman2017, ofmanModelingIonBeams2022, pezziniFullyKineticSimulations2024, verniero2020, verniero2022, wu2025}. Because collisions are weak in the fast SW, kinetic microinstabilities play a crucial role in regulating these deviations from equilibrium \citep{garyTheorySpacePlasma1993, gary2000, gary2003, hollweg2014, lu2006}. In-situ measurements have demonstrated that the SW is confined to regions of parameter space bounded by the thresholds of various instabilities \citep{bale2009, bourouaine2013, hellinger2006, hellinger2011, kasper2002, marsch2004, maruca2012, matteini2007}. When an instability threshold is exceeded, the resulting instability acts to reduce deviations from thermodynamic equilibrium by generating plasma waves. These waves interact with particles and reshape their distribution functions through various mechanisms including: cyclotron resonance \citep{araneda2008, araneda2009, gary1999, goldstein1994, hollweg2002, leamon1998,  marsch2001, matteini2007}, as well as stochastic heating by oblique kinetic Alfv\'{e}n waves via Landau damping \citep{leamon1998, leamon1999, leamon2000, howes2008}.

Observations of the fast SW indicate that the absolute value of the typical relative velocity between alpha particles and protons is typically comparable to the local Alfv\'{e}n speed $c_{A p} \doteq \|\boldsymbol{B}\| / (4\pi \rho_p)^{1/2}$, where $\rho_p$ is the proton mass density, which decreases with distance from the Sun \citep{marsch1982b, reisenfeld2001, verscharen2015}. Previous studies have shown that this relative drift constitutes a reservoir of free energy, and its release through plasma instabilities in the form of waves contributes significantly to SW heating, simultaneously causing a continuous deceleration of the alpha particles \citep{borovsky2014, feldman1979, safrankova2013, schwartz1981}. When the drift velocity exceeds the local Alfv\'{e}n speed, it can excite the fast-magnetosonic/whistler (FM/W) instability \citep{gary2000, li2000, revathy1978} as well as the Alfv\'{e}n/ion-cyclotron (A/IC) instability \citep{martinovic2025, verscharen2013a}. According to the linear theory (LT), these waves can interact resonantly with protons, producing a secondary proton beam \citep{tu2002}. The discrimination of different ion species in satellite data remains a complex challenge. Nevertheless, recent technical advances in ion component separation from satellite observations \citep{demarco2023} have provided new insights, moving us closer to identifying the origin of alpha particle beams, which may not be generated locally \citep{bruno2024}. A clear consensus has not yet been reached, underscoring the need for continued investigation. 

In this work, we propose a local mechanism for the generation of alpha particles via non-linear Landau damping. The remainder of this paper is organized as follows. In Section~\ref{sec:setup}, we present the numerical setup of the particle-in-cell (PIC) simulation, defining the initial conditions and outlining the hypotheses under which the system is modeled. We describe both the kinetic and numerical parameters, the latter chosen to ensure stability and an accurate resolution of the physical quantities. In Section~\ref{sec:results}, we present the results in the following order. Subsection~\ref{subsec:linear} discusses the LT of the unstable eigenmodes developing from the initial conditions, with particular emphasis on the choice of the reduced mass ratio and how the wave might dissipate. Subsection~\ref{subsec:pic} reports the findings from the fully kinetic PIC simulation, including the system energetics, spectral analysis, and VDFs. Subsection~\ref{subsec:fpc} focuses on the analysis of field-particle interactions from the simulation data. In Section~\ref{sec:disc}, we provide a comprehensive discussion comparing the LT results with those from the PIC simulations. To conclude, Section~\ref{sec:concl} summarises our findings and places them in the context of the established literature.

%–––––––––––––––––––––––––––––––––––––––––––––––––––––––––––––––––––%
%                      NUMERICAL SET-UP SECTION                     %
%–––––––––––––––––––––––––––––––––––––––––––––––––––––––––––––––––––%

\section{Numerical setup} \label{sec:setup}

We use the \textsc{ECsim} PIC code \citep{lapentaExactlyEnergyConserving2017,gonzalez-herreroECsimCYLEnergyConserving2019,bacchiniRelSIMRelativisticSemiimplicit2023, croonen2024} to solve the Vlasov-Maxwell system for a low-$\beta$ electron-proton-alpha plasma in the Newtonian regime. The simulation is performed in a one-dimensional periodic Cartesian domain, neglecting expansion, since $\tau_{\text{exp}} / \tau_{\text{lin}} \sim 10^{3}$ expansion versus linear instability temporal scale ratio. Therefore, there is negligible interplay between the two processes. The expansion timescale is estimated using the approximation $\tau_{\text{exp}} \approx R_{\odot} / v_{\text{sw}}$, assuming an Alfv\'enic wind ($v_\mathrm{sw}\sim v_{Ap}$), while $\tau_{\text{lin}}\doteq 2\pi/\gamma_\text{max}$ is the maximum linear instability time scale and $\gamma_\text{max}$ its growth rate. Hereafter, quantities with a ``0'' in the subscript refer to their values at the initial time. We adopt the de Hoffmann-Teller frame to eliminate the initial electric field $\boldsymbol{E}_0 =\boldsymbol{0}$, while the background magnetic field, which is invariant under Galilean frame transformations, is uniform and directed along the $x$-axis, i.e., $\boldsymbol{B}_0 = B_0 \hat{\boldsymbol{e}}_x$. In the magnetic-field-aligned frame, the $x$-axis therefore defines the parallel direction, while the $y$- and $z$-axes are considered perpendicular. From this point onward, all quantities with parallel or perpendicular subscripts are implicitly expressed in the field-aligned coordinate system.

% Schematic cartoon of the simulation setup
\begin{figure*}[ht!]
\centering
%trim={<left> <bottom> <right> <top>}
%You must also include the clip option, otherwise the image will still occupy the original space.
\includegraphics[width=\linewidth,trim={0cm 3cm 0cm 3cm},clip]{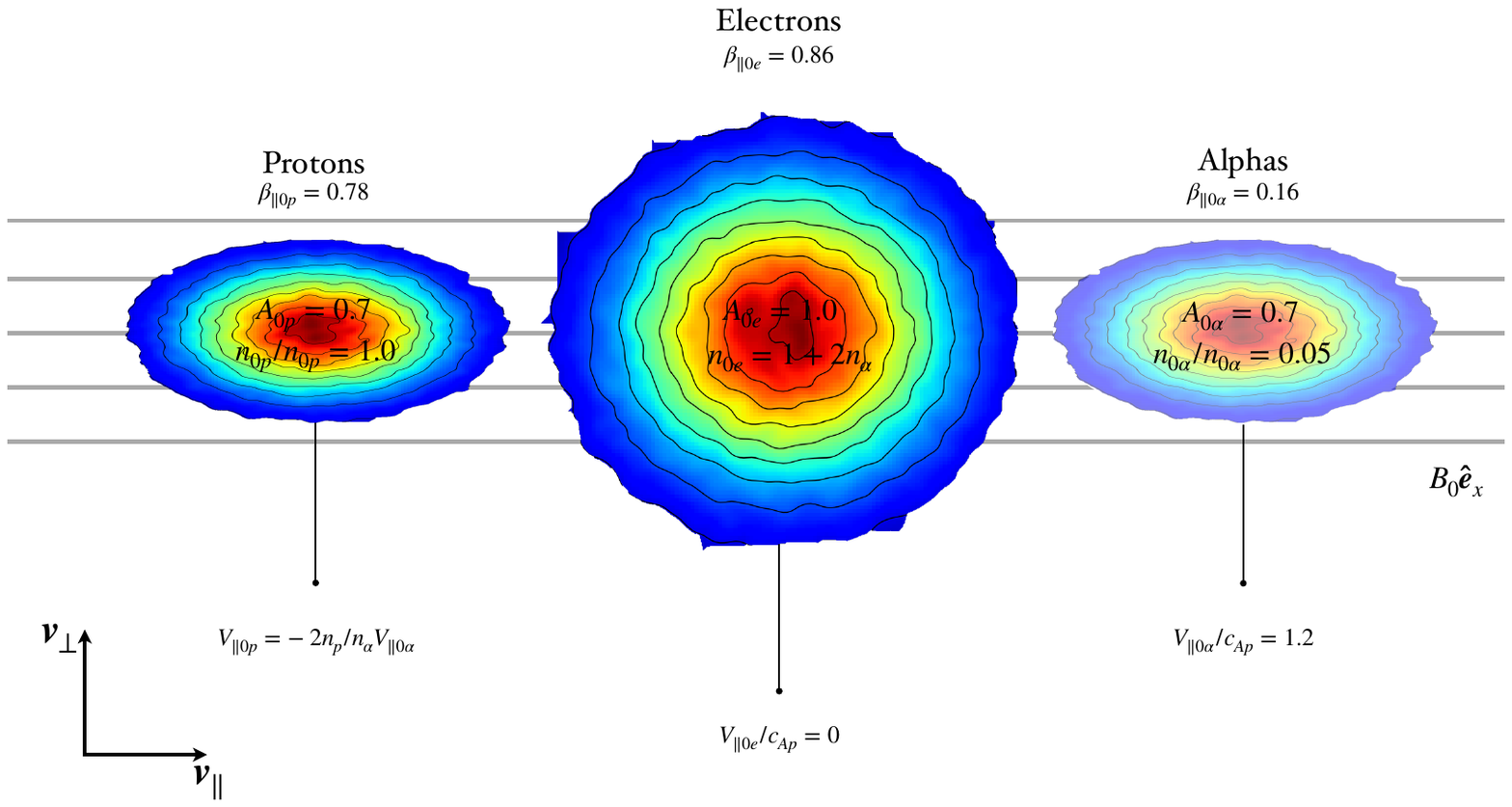}
\caption{Schematic illustration of the simulation setup. The figure shows a portion of velocity phase space containing the electron and ion VDFs, with color intensity indicating particle density. For each species, the plasma-$\beta$ parameter is defined as $\beta_s \doteq 8\pi n_s k_B T_s / B^2$.}
\label{fig:sketch}
\end{figure*}

Electrons are in Maxwellian equilibrium at initialization, therefore their drift (bulk) velocity $V_{\| 0 e}=0$ and their thermal speed is isotropic. For each particle species $s$, the drift speed is defined as $\boldsymbol{V}_{s} \doteq \int \mathrm{d}^3 \boldsymbol{v} \boldsymbol{v} f_{s}/n_{s}$, where $n_s$ the number density and $f_{s}$ is the distribution function. The ion species, protons and alphas, are initialized with a bi-Maxwellian distribution 
%\onecolumngrid
 \begin{equation}
        \begin{aligned}
            f_{\textrm{bi-M}, s}\left(v_{\|}, v_{\perp}\right)\doteq &  \frac{n_{0 s}}{\sqrt{\pi} v_{\| 0 s}} \exp \left[-\frac{\left(v_{\|}-V_{\| 0 s}\right)^2}{v_{\| 0 s}^2}\right] \\
            & \times \frac{1}{\pi v_{\perp 0 s}^2} \exp \left(-\frac{v_{\perp}^2}{v_{\perp 0 s}^2}\right),
        \end{aligned}
        \label{eq:maxwell}
\end{equation}
such that each species is characterized by a non-zero drift velocity along the parallel direction and distinct thermal speeds.
% \twocolumngrid

\begin{table}[ht!]
\centering
\caption{Initial simulation parameters. \label{tab:kinetic}}
\begin{tabular}{lcccc}
\hline \hline
Parameters & Electrons &  Protons & alphas \\
\hline
$n_{0 s}/n_{0 p}$ & $1.10$ & $1.00$ & $5.00 \times 10^{-2}$ \\
$v_{\| 0 s}/c_{A p}$ & $6.25$ & $6.25\times 10^{-1}$ & $6.25$ \\
$v_{\perp 0 s}/c_{A p}$ & $6.25$ & $4.42 \times 10^{-1}$ & $4.42 \times 10^{-1}$ \\
$V_{\| 0 s}/c_{A p}$ & $0$    & $- 0.12 $ & $1.2$ \\
\hline
\end{tabular}
\end{table}

The simulation is initialized by inserting into Equation~\eqref{eq:maxwell} the parameters listed in Table~\ref{tab:kinetic}. We provide a visual representation of the initial state of all particle species in Figure~\ref{fig:sketch}. Here we define the thermal speed in the direction $j = \parallel, \perp$ as $v_{js} = (k_B T_{js} / m_s)^{1/2}$, where $k_B$ is Boltzmann's constant, $m_s$ the mass of species $s$, and $T_{js}$ the temperature in the direction $j$ of species $s$.
%The schematic cartoon in Figure~\ref{fig:sketch} corresponds to this initialization and provides a visual representation of the velocity phase space as defined by the distribution function $f_{\textrm{bi-M}, s}$ previously introduced.
The parameters in Table~\ref{tab:kinetic} are chosen to satisfy the conditions of charge and current neutrality, as follows:

 \begin{equation}
        \begin{aligned}
            \begin{cases}
            \sum_{s} Z_{s} q_{s} n_{0s} = 0
            & \\ 
            \sum_{s} Z_{s} q_{s} n_{0 s} V_{\| 0s}  = 0,
            \end{cases}
        \end{aligned}
        \label{eq:nutrality}
\end{equation}
where $Z$ is the atomic number and $q$ is the elementary charge. Since alpha particles drift relatively fast with respect to a rest frame, we expect a beam-plasma interaction driven by alphas.

We set the domain length in the parallel direction to $L_{x}/d_{p} = 160.0$ in order to properly accommodate multiple oscillations of the fastest growing eigenmode (see Section~\ref{subsec:linear}). Here, $d_{p} \doteq c/\omega_{pp}$ is the proton skin depth, $\omega_{pp}\doteq (4\pi e^2 n_p /m_p)^{1/2}$ is the proton plasma frequency, $m_{p}$ the mass and density of protons and $c$ is the speed of light, and in our normalization choice $\omega_{pp} = c =1$. Setting the number of grid cells in the parallel direction to $N_{ x} = 1024$ gives a spatial resolution of $\Delta x = L_{x}/N_{x} = 0.15625 d_p$. This resolution more than adequately resolves the proton and alpha cyclotron radius, which for the ``$s$'' species is defined as $\rho_{s}\doteq v_{s}/\Omega_{cs}$, with $v_{s}$ the magnitude of the thermal speed and $\Omega_{cs}\doteq Z_{s}e B_0/m_{s} c$ the cyclotron frequency. This results in $\rho_{p}/\Delta x = 4$ and $\rho_{\alpha}/\Delta x = 8$. Electrons are slightly underresolved, with $\rho_{e}/\Delta x \approx 0.5$, due to our choice of mass ratio $m_{p}/m_{e} = 100$. Restricting the analysis to the parallel direction allows us to focus exclusively on quasi-parallel modes while neglecting those in the perpendicular direction. 
%This is expected since the electron cyclotron radius scales with the square root of the mass ratio, i.e., as $m_{p}/m_{e}^{1/2}$. In this way, the electron gyration is less under-resolved than it would be with a realistic mass ratio.

Similarly, the temporal domain $t\in[0, 50 \tau_{\text{lin}})$ is chosen to span several times the characteristic time scale of the fastest growing mode. The temporal resolution is set to $\Delta t/\omega_{pp}^{-1} = 0.078$, which allows us to resolve the plasma periods of all species $\Pi_{ps} \doteq 2\pi/\omega_{ps}$. Specifically, we have $\Pi_{pe}/\Delta t \approx 4$, $\Pi_{pp}/\Delta t \approx 40$, and $\Pi_{p\alpha}/\Delta t \approx 180$ which abundantly resolve the gyroperiod of ions and electrons. 
%
%These choices satisfy the code’s stability condition for avoiding the finite grid instability: $v_{\text{max}} \Delta t / \Delta x < 1$.
%
We employed $4096$ particles per cell per species, initially distributed uniformly on the grid.

%–––––––––––––––––––––––––––––––––––––––––––––––––––––––––––––––––––%
%                          RESULTS SECTION                          %
%–––––––––––––––––––––––––––––––––––––––––––––––––––––––––––––––––––%

\section{Results}\label{sec:results}

%–––––––––––––––––––––––––––––––––––––––––––––––––––––––––––––––––––%
%                  LINEAR THEORY SUB-SECTION                %
%–––––––––––––––––––––––––––––––––––––––––––––––––––––––––––––––––––%

\subsection{Linear Theory}\label{subsec:linear}
% Introduction
The LT of the Vlasov--Maxwell system, with our choice of initial conditions, provides the solutions of the dispersion relation, which serves as a first-order approximation of the non-equilibrium plasma. In this work, we employ the \textsc{DIS-K}\footnote{The code is publicly available at \url{https://github.com/ralopezh/dis-k}.} linear solver \citep{lopezGeneralDispersionProperties2021,lopezRalopezhDiskFirst2023} to: (i) determine the optimal parameter setup for the non-linear fully kinetic simulation (see Section~\ref{sec:setup}); and (ii) assess the consistency of the fully kinetic simulation within its linear phase in comparison with the linear solution (see Section~\ref{sec:results}). In particular, comparing the growth rate of the instability obtained from the LT with that from the non-linear fully kinetic simulation during its linear phase allows us to confirm the validity and quantitative reliability of the simulation, thereby ensuring the physical relevance of the results.

% Linear theory k_parallel vs k_perp 
\begin{figure*}[ht!]
\includegraphics[width=1\textwidth,trim={1cm 0 2.5cm 0},clip]{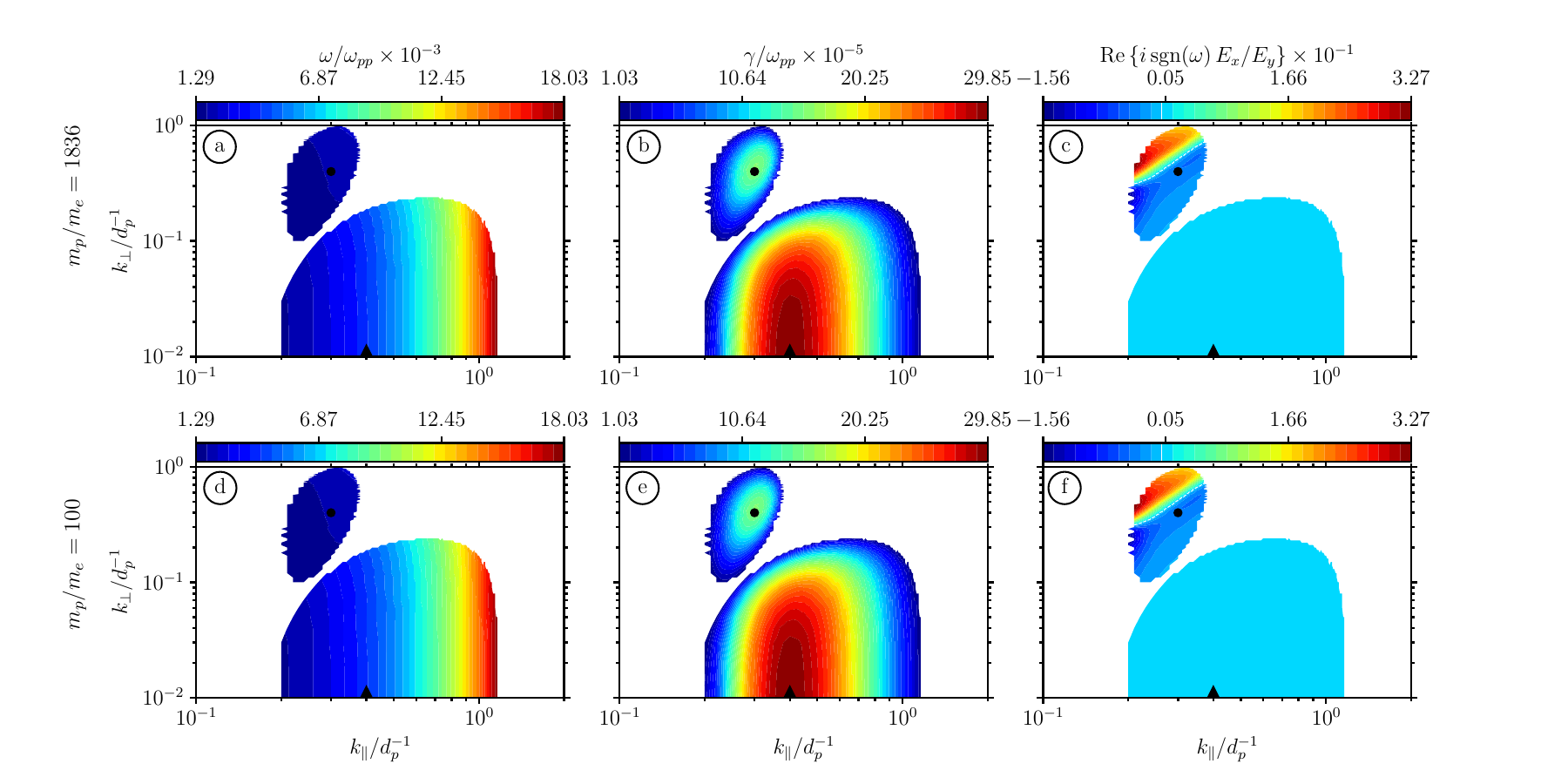}
\caption{In the first row, panels~(a) to (c) respectively show the result of LT for the real frequency $\omega_{r}/\omega_{pp}$, growth rate $\gamma/\omega_{pp}$, and polarization of the unstable eigenmode in the $k_{\parallel}$-$k_{\perp}$ plane,
%which is respectively the space defined by the wavenumber in the parallel and perpendicular direction with respect to the magnetic field,
computed using the realistic mass ratio $m_p/m_e = 1836$.
%Here, $\omega_{pp}\doteq \sqrt{4\pi e^2 n_p /m_p}$ is the proton plasma frequency, $e$ is the fundamental charge, $m_{p}$ and $n_{p}$ are respectively the mass and density of protons.
In the second row, panels~(d) to (f), the same quantities are shown but computed with a reduced mass ratio of $m_p/m_e = 100$. The black triangle indicates the most unstable quasi-parallel propagating eigenmode, while the black dot indicates the most unstable oblique propagating eigenmode.  In panels~(c) and (f), the white dotted line indicates the contour separating right-hand (RH) from left-hand (LH) polarization.}
\label{fig:qlt}
\end{figure*}

In Figure~\ref{fig:qlt}, the unstable eigenmode spectrum from LT is plotted in the $k_{\parallel}$-$k_{\perp}$ plane, showing specifically: real wave frequency $\omega/\omega_{pp}$ in panels~\ref{fig:qlt}(a) and (d), instability growth rate $\gamma/\omega_{pp}$ in panels~\ref{fig:qlt}(b) and (e), and wave polarization in panels~\ref{fig:qlt}(c) and (f). We compare the spectrum obtained with a realistic mass ratio of $m_p/m_e = 1836$ in the top row with the spectrum computed with a reduced mass ratio of $m_p/m_e = 100$ in the bottom row. The spectrum does not present any substantial differences, both qualitatively and quantitatively, with respect to the variation of the proton-to-electron mass ratio. The most unstable eigenmode, for both the quasi-parallel and oblique propagating eigenmode branches, appears morphologically identical for both mass ratios. For this reason, from now on, we will refer only to the calculation with $m_p/m_e = 100$. In order to distinguish the different unstable branches, we introduce the branch index ``$i$'': when $i = 1$, we refer to the quasi-parallel propagating eigenmode branch (indicated with the black triangle in Figure~\ref{fig:qlt}) $(k_{\parallel}, k_{\perp})_{1}^{\star}/d_{p}^{-1} \approx (0.4, 0.01)$, with quasi-parallel direction of propagation $\theta \lesssim 10^{\circ}$. When $i = 2$, we refer to the oblique propagating eigenmode branch (indicated by the black dot in the same figure) $(k_{\parallel}, k_{\perp})_{2}^{\star}/d_{p}^{-1} \approx (0.3, 0.35)$. Panel~\ref{fig:qlt}(d) shows that the maximum value of the wave frequency is positive, $\omega_{i}^{\star}/\omega_{pp} \in \mathrm{Re}_{+}$, for both branches, with values $\omega_{1}^{\star}/\omega_{pp}\approx 4.12 \times 10^{-3}$ and $\omega_{2}^{\star}/\omega_{pp}\approx 1.29\times 10^{-3}$, meaning that the two eigenmodes are non-stationary with a positive direction of propagation. Panel~\ref{fig:qlt}(e) shows the eigenmode growth rate $\gamma/\omega_{pp}$, with maximum values $\gamma_{1}^{\star}/\omega_{pp} \approx 2.99 \times 10^{-4}$ and $\gamma_{2}^{\star}/\omega_{pp} \approx 1.52 \times 10^{-4}$, confirming that the system supports two unstable eigenmodes amplified on linear timescales $\tau_{\mathrm{lin}, i} \doteq 2\pi/\gamma_{i}^{\star}$. Panels~\ref{fig:qlt}(c) and (f) show the unstable eigenmode polarization $P\doteq \mathrm{Re}\left\{ i\,\mathrm{sgn}(\omega)\, E_x/E_y \right\}$ \citep{garyTheorySpacePlasma1993, stixWavesPlasmas1962}, where $E_x$ and $E_y$ are the electric field components in the $x$ and $y$-direction. Here, $P \in \mathrm{Re}_{+}$ means that the wave is right-handed (RH) for both branches, i.e.\ it rotates counter-clockwise in a plane perpendicular to the direction of propagation; the wave is also circularly polarized (CP) for the quasi-parallel eigenmode branch since $P_1^{\star}\approx 1$, and elliptically polarized (EP) for the oblique eigenmode branch since $P_2^{\star}\lesssim 1$.

% Linear theory in k-theta plane
\begin{figure}
\centering
\includegraphics[width=\columnwidth]{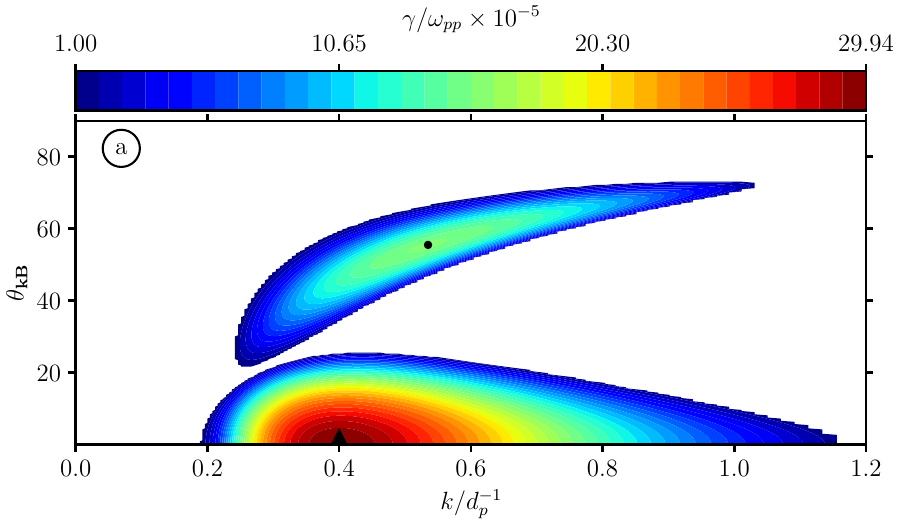}
\caption{Growth rate plotted in the ${k}$-$\theta_{\boldsymbol{k}\boldsymbol{B}}$ plane computed using the reduced proton-to-electron mass ratio. The black triangle marks the parallel propagating unstable eigenmode, while the black dot marks the oblique propagating one.}
\label{fig:ktheta}
\end{figure}

Figure~\ref{fig:ktheta} shows the growth rate of the unstable eigenmode plotted in the ${k}$-$\theta_{\boldsymbol{k}\boldsymbol{B}}$ plane, where $\theta_{\boldsymbol{k}\boldsymbol{B}}$ is the angle between the wavevector $\boldsymbol{k}$ and $\boldsymbol{B}$. In this plane, it is easier to distinguish the origin of the two main unstable branches. The black triangle indicates the quasi-parallel propagating eigenmode, whose main centroid indicates a maximum growth rate located at $\theta_{\boldsymbol{k}\boldsymbol{B}}^{\star} \lesssim 5^{\circ}$; these eigenmodes correspond to the FM/W branch. The black dot marks the oblique propagating eigenmodes, whose centroid indicates a maximum growth rate located at $\theta_{\boldsymbol{k}\boldsymbol{B}}^{\star} \approx 55^{\circ}$, corresponding to the A/IC branch.

To perform an eigenmode stability analysis, it is necessary to consider not only the amplification of eigenmodes driven by the alpha-to-proton relative drift and the thermal anisotropy, but also the possible damping effects caused by the plasma species that constitute the system, in this case protons and alpha particles, as discussed in \cite{verscharen2013}. A particle can interact strongly with an electromagnetic wave Doppler-shifted with respect to its cyclotron frequency or its harmonics, exchanging energy and momentum \citep[e.g.,][]{narita2017}; the condition for wave particle resonant damping can be written as
\begin{equation}
    \omega = \boldsymbol{k} \cdot \boldsymbol{v} + n \Omega_{c}.
    \label{eq:resonance}
\end{equation}
Here, $\Omega_{c}$ is the cyclotron frequency and the resonance index $n\in \mathbb{Z}$ refers to the harmonics of the wave. Restricting the motion in the direction parallel to the magnetic field, the wavenumber becomes $\boldsymbol{k} = k_{\parallel} \hat{\boldsymbol{b}}$ and the particle speed $\boldsymbol{v} = v_{\parallel} \hat{\boldsymbol{b}}$, where $\hat{\boldsymbol{b}} \doteq \boldsymbol{B}/\|\boldsymbol{B}\|$. In this way, particles can exchange energy with the wave's electric field both along the mean magnetic field (Landau damping) and perpendicularly to it (cyclotron damping) \citep{narita2017, tsurutaniBasicConceptsWaveparticle1997, verscharen2013}. 
The so-called ``normal'' first-order cyclotron damping ($n \in \mathbb{Z}_{>0}$) can be represented as a head-on collision between a charged particle and a wave with identical polarization, in which the relative motion between the wave and the particle leads to a Doppler shift of the wave frequency up to the cyclotron frequency $\omega - k_{\|} v_{\|} = +\Omega_c$. An example is the interaction between an ion moving parallel to the magnetic field in the positive direction ($\boldsymbol{v}_{i} \cdot \boldsymbol{B} > 0$; positively charged particles undergo clockwise cyclotron motion under the Lorentz force excited by the magnetic field) and a left-hand circularly polarized (LHCP) wave propagating antiparallel to the magnetic field direction ($\boldsymbol{v}_{\text{ph}} \cdot \boldsymbol{B} < 0$, where $v_{\text{ph}}\doteq \omega / k_{\parallel}$ is the phase speed of the wave). In this configuration, the ion gyration and the wave's polarization\footnote{\cite{stixWavesPlasmas1962} defines right- or left-handed polarization as the sense of rotation, in time, of a fluctuating field vector at a fixed point in space, when viewed along the direction of the background magnetic field $\boldsymbol{B}_0$ at a positive real frequency. Under this definition, a right-hand eigenmode propagating either parallel or anti-parallel to $\boldsymbol{B}_0$ has fluctuating field vectors that rotate in the same sense as the cyclotron motion of an electron with $v_z = 0$. This eigenmode corresponds to the FM/W in a stable plasma under parallel propagation. Similarly, a left-hand eigenmode rotates in the same sense as the cyclotron motion of an ion and corresponds to the A/IC wave in a stable plasma with parallel propagation.} are chiral
%\footnote{An object or system is called chiral if it cannot be superimposed onto its mirror image by any combination of rotations and translations. In other words, chiral objects have a ``handedness'' (like left and right hands) and exhibit asymmetry with respect to their mirror image.} 
and thus resonant, enabling energy exchange through cyclotron damping. Similarly, electrons moving parallel to the magnetic field ($\boldsymbol{v}_{e} \cdot \boldsymbol{B} > 0$), which exhibit counterclockwise cyclotron motion with respect to the local magnetic field, can interact resonantly with a right-hand circularly polarized (RHCP) wave propagating antiparallel to the magnetic field direction ($\boldsymbol{v}_{ph} \cdot \boldsymbol{B} < 0$). 

%Another type of resonance is the so-called ``anomalous'' cyclotron resonant damping ($n \in \mathbb{Z}_{<0}$), which can be represented as a tail-on collision between a charged particle and a wave with anti-symmetric polarization. An example is the interaction between an ion moving parallel to the magnetic field in the positive direction ($\boldsymbol{v}_{i} \cdot \boldsymbol{B} > 0$) and an RHCP wave also propagating parallel to the magnetic field ($\boldsymbol{v}_{ph} \cdot \boldsymbol{B} > 0$). In this configuration, the ion gyration and the wave's polarization are anti-symmetric. The left-hand-rotating ion overtakes the right-hand wave ($\|\boldsymbol{v}_{i}\| > \|\boldsymbol{v}_{ph}\|$) and perceives it as left-hand polarized in its own frame. The Doppler shift decreases the wave frequency to that of the cyclotron frequency, $\omega - k_{\|} v_{\|} = -\Omega_c$: the relative motion between the particle and the wave Doppler-shifts the wave frequency down to the ion cyclotron frequency. This interaction is called ``anomalous'' because RH waves, which normally interact with RH (electron-like) cyclotron motion, here resonate with LH (ion-like) cyclotron motion instead. A similar argument applies for electrons interacting with LH polarized waves.

Another form of resonance is the so-called ``anomalous'' cyclotron resonant damping ($n \in \mathbb{Z}_{<0}$), which can be described as a tail-on encounter between a charged particle and a wave with opposite-sense polarization. For instance, consider the interaction between an ion traveling parallel to the magnetic field in the positive direction ($\boldsymbol{v}_{i} \cdot \boldsymbol{B} > 0$) and an RHCP wave also moving along the magnetic field ($\boldsymbol{v}_{ph} \cdot \boldsymbol{B} > 0$). In this situation, the ion gyration and the wave polarization are oppositely oriented. The left-hand-rotating ion overtakes the right-hand wave ($\|\boldsymbol{v}_{i}\| > \|\boldsymbol{v}_{ph}\|$) and perceives it as left-hand polarized in its own frame. Due to the Doppler effect, the wave frequency is shifted down to match the cyclotron frequency, $\omega - k_{\|} v_{\|} = -\Omega_c$: the relative motion between the ion and the wave causes the frequency to be Doppler-shifted to the ion cyclotron frequency. This process is termed ``anomalous'' because RH waves, which usually resonate with RH (electron-like) cyclotron motion, here interact with LH (ion-like) cyclotron motion instead. An analogous reasoning holds for electrons interacting with LH-polarized waves.

In conclusion, for a generic species ``$s$'', cyclotron damping occurs at wavenumbers and frequencies that satisfy the condition
\begin{equation}
    -k_{\parallel} v_{\| s} + \Omega_{cs} \lesssim \omega \lesssim k_{\parallel} v_{\| s} + \Omega_{cs}.
    \label{eq:thresholdn1}
\end{equation}
Similarly, Landau damping (with $n = 0$) occurs at wavenumbers and frequencies that fulfill
\begin{equation}
    -k_{\parallel} v_{\|s} \lesssim \omega \lesssim k_{\parallel} v_{\| s}.
    \label{eq:thresholdn0}
\end{equation}
alpha cyclotron resonance driving (with $n = -1$) occurs at wavenumbers and frequencies satisfying
\begin{equation}
    \omega =  (V_{\alpha} + \sigma v_{\| \alpha}) k_{\parallel} - \Omega_{c \alpha},
    \label{eq:alpharesonance}
\end{equation}
where $\sigma$ is a parameter defined empirically in \cite{verscharen2013a}, which takes values between $[1, 3)$. In our case, we considered the lower limit, fixing $\sigma=1$.

% Linear theory dispersion relation and resonance cone
\begin{figure}[ht!]
\centering
\includegraphics[width=\columnwidth]{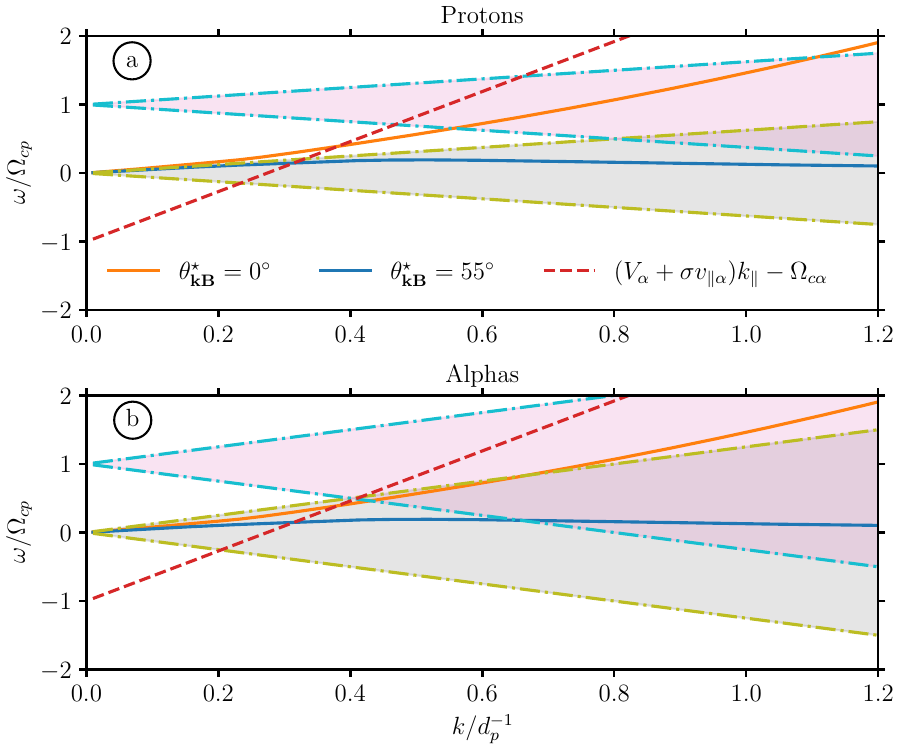}
\caption{Dispersion relation and resonance conditions for the quasi-parallel FM/W eigenmode with $\theta_{\boldsymbol{k}\boldsymbol{B}}^{\star} = 0^{\circ}$ (orange solid line) and for the oblique A/IC eigenmode with $\theta_{\boldsymbol{k}\boldsymbol{B}}^{\star} = 55^{\circ}$ (blue solid line). The dashed red line represents the $n = -1$ resonance condition for as, given by Equation~\eqref{eq:alpharesonance}. Panel~(a) shows quantities related to protons: the cyan dot-dashed lines, defined by Equation~\eqref{eq:thresholdn1}, indicate the thresholds for cyclotron resonance, while the green lines, defined by Equation~\eqref{eq:thresholdn0}, indicate those for Landau resonance. Panel~(b) shows the corresponding quantities for alphas.
}
\label{fig:disp}
\end{figure}

In Figure~\ref{fig:disp}, panels~(a) and (b) show the LT dispersion relation for the unstable FM/W eigenmode (orange solid line), restricted to purely parallel propagation with $\theta_{\boldsymbol{k}\boldsymbol{B}}^{\star} = 0^\circ$, and for the unstable A/IC eigenmode (blue solid line), restricted to oblique propagation with $\theta_{\boldsymbol{k}\boldsymbol{B}}^{\star} = 55^\circ$. In both panels, the red dashed line represents the cyclotron resonance with index $n = -1$, evaluated with alpha-particle parameters.

Panel~\ref{fig:disp}(a) includes the potential effect of wave dissipation through proton resonance. The cyan dash-dotted line, representing Equation~\eqref{eq:thresholdn1} evaluated with proton parameters, and the green dash-dotted line, representing Equation~\eqref{eq:thresholdn0}, define the regions where unstable waves can be absorbed by protons through cyclotron ($n = +1$) and Landau ($n = 0$) resonances, respectively. The area between the purple lines (light gray region) indicates the Landau-resonant region, while the area between the green lines (light pink region) indicates the cyclotron-resonant region.
The dispersion relation associated with the oblique-propagating eigenmode at $\theta_{\boldsymbol{k}\boldsymbol{B}}^{\star} = 55^\circ$ lies entirely within the pink region, suggesting that dissipation via proton Landau resonance is possible. In contrast, the parallel-propagating eigenmode at $\theta_{\boldsymbol{k}\boldsymbol{B}}^{\star} = 0^\circ$ intersects the cyclotron threshold in the range $0.6 \lesssim k/d_p^{-1} \lesssim 1.9$, but not the Landau threshold. Therefore, this eigenmode may resonate with alpha particles where the red dashed line intersects the orange solid line. From the perspective of alpha cyclotron resonance, the gray and pink areas can be considered ``forbidden'' regions where proton absorption dominates.

Panel~\ref{fig:disp}(b) includes the potential effect of wave dissipation through alpha particle resonance. The purple dash-dotted line, representing Equation~\eqref{eq:thresholdn1} evaluated with alpha parameters, and the green dash-dotted line, representing Equation~\eqref{eq:thresholdn0}, define the regions where unstable waves can be absorbed by alphas via cyclotron ($n = +1$) and Landau ($n = 0$) resonances, respectively. The same considerations discussed above for protons also apply here to alphas.

The dispersion relation associated with the oblique-propagating eigenmode at $\theta_{\boldsymbol{k}\boldsymbol{B}}^{\star} = 55^\circ$ lies entirely within the gray region, suggesting that this eigenmode could be dissipated through Landau resonance with alpha particles. Similarly, the parallel-propagating eigenmode at $\theta_{\boldsymbol{k}\boldsymbol{B}}^{\star} = 0^\circ$ intersects the Landau threshold in the range $0 \lesssim k/d_p^{-1} \lesssim 0.7$ and the cyclotron threshold in the range $0.4 \lesssim k/d_p^{-1} \lesssim 1.2$. Therefore, cyclotron resonance with $n = -1$ may compete with Landau resonance, since the intersection between the $\theta_{\boldsymbol{k}\boldsymbol{B}}^{\star} = 0^\circ$ dispersion curve and the red dashed line lies within the gray region. In this case, the forbidden regions (gray and pink) cover the entire range of unstable frequencies, potentially suppressing $n=-1$ resonance through resonant damping ($n=0, +1$) by alphas. 

%–––––––––––––––––––––––––––––––––––––––––––––––––––––––––––––––––––%
%                  LINEAR THEORY SUB-SECTION                %
%–––––––––––––––––––––––––––––––––––––––––––––––––––––––––––––––––––%

\subsection{Fully Kinetic Simulation}\label{subsec:pic}
\begin{figure}[ht!]
\centering
\includegraphics[width=\columnwidth]{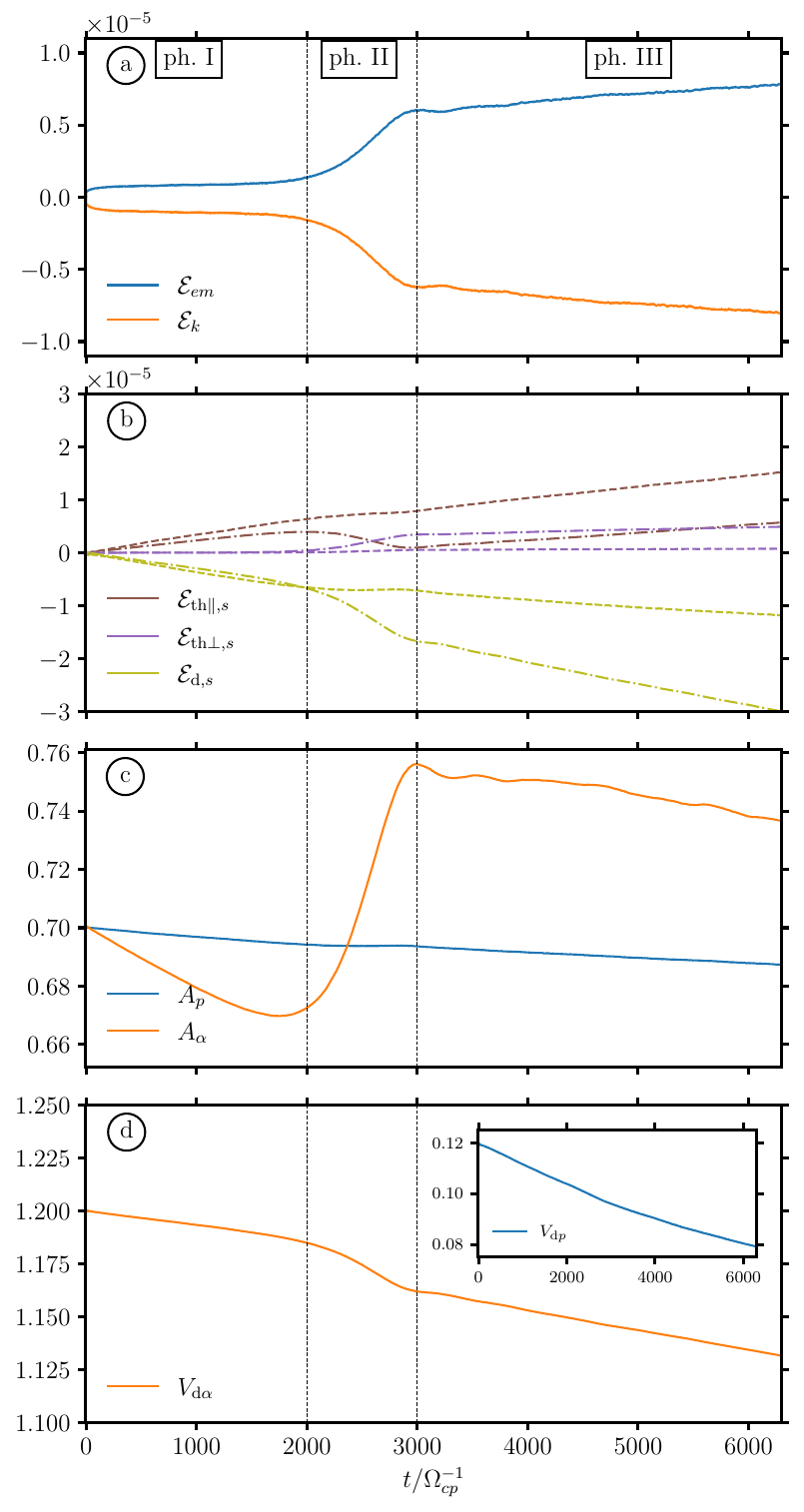}
\caption{Time series of relevant kinetic quantities are shown, indicating the three distinct temporal phases of the system’s evolution (vertical dashed lines). These quantities are defined in Section \ref{sec:results}. Panel (a): system’s global electromagnetic energy (blue solid line) and global kinetic energy (orange solid line). Panel (b): thermal energy in the parallel direction (brown line), thermal energy in the perpendicular direction (cyan line) and the drift energy (green line), respectively, for protons (dashed lines) and alpha particles (dash–dotted lines). Panel (c): temporal evolution of ion anisotropy; panel (d): temporal evolution of ion drift speed.}
\label{fig:energy}
\end{figure}

The time series of the most relevant kinetic quantities are shown in Figure~\ref{fig:energy}, to provide an overview of the system’s global evolution. In panels~\ref{fig:energy}(a) and (b), all the plotted energies $\mathcal{E}$ represent the relative change with respect to their initial values $E_{0}$, normalized to the system’s total initial energy $U_{0}$, as $\mathcal{E} = (E - E_{0})/U_{0}$. The system’s electromagnetic energy $\mathcal{E}_{em}$ and kinetic energy $\mathcal{E}_{k}$, defined as

\begin{equation}
    \begin{aligned}
        \mathcal{E}_{em} &\doteq \int \mathrm{d}\boldsymbol{x}\, \frac{B^2 + E^2}{8\pi} \\
        \mathcal{E}_{k} &\doteq \sum_{s} \left(\mathcal{E}_{th, s}+ \mathcal{E}_{\textrm{d},s}\right)
    \end{aligned}
    \label{eq:kinmag}
\end{equation}
Here, we define the components of $\mathcal{E}_{k}$ as follows:
\begin{equation}
    \begin{aligned}
        \mathcal{E}_{th, s} &\doteq \mathcal{E}_{th\perp, s}+ \mathcal{E}_{th\|, s}= \int \mathrm{d}\boldsymbol{x} \left(p_{\perp,s} + \frac{p_{\parallel,s}}{2} \right)\\
        \mathcal{E}_{\textrm{d},s} &\doteq \int \mathrm{d}\boldsymbol{x}\frac{\rho_{s} V_{s}^2 }{2}.
    \end{aligned}
    \label{eq:kinspec}
\end{equation}
% Result 1: Energies temporal series of the system

In equation~\ref{eq:presstensor} we properly define the pressure tensor and its perpendicular and parallel component for the generic species $s$.

\begin{equation}
\begin{aligned}
    \boldsymbol{P}_{s} &\doteq m_{s} \int \mathrm{d}^3 \boldsymbol{v} (\boldsymbol{v} - \boldsymbol{V}_{s})(\boldsymbol{v} - \boldsymbol{V}_{s}) f_{s}, \\
    p_{\perp, s} &\doteq \boldsymbol{P}_{s} \bdbldot (\mathbb{1} - \hat{\boldsymbol{b}} \hat{\boldsymbol{b}})/2, \\
    p_{\parallel, s} &\doteq \boldsymbol{P}_{s} \bdbldot \hat{\boldsymbol{b}} \hat{\boldsymbol{b}}.
\end{aligned}
\label{eq:presstensor}
\end{equation}

The time evolution of the two energies in Equation~\eqref{eq:kinspec}, shown in panel~\ref{fig:energy}(a), illustrates the conversion of $\mathcal{E}_{k}$ into $\mathcal{E}_{em}$ as a consequence of the system’s instability. The system exhibits three distinct phases, indicated in the plots by black dashed vertical lines:
\begin{itemize}
    \item \textbf{Excitation phase (I)}--$0 \lesssim t/\Omega_{cp}^{-1} \lesssim 2{,}000$: The system remains in a metastable equilibrium, perturbed only by the numerical noise inherent in full-kinetic algorithms. The system remains in this state until the perturbations overcome the potential barrier, triggering the instability.
    
    \item \textbf{Growth phase (II)}--$2{,}000 \lesssim t/\Omega_{cp}^{-1} \lesssim 3{,}000$: The instability develops; $\mathcal{E}_{em}$ rises rapidly at the expense of $\mathcal{E}_{k}$, which decreases correspondingly.
    
    \item \textbf{Secular growth phase (III)}--$t/\Omega_{cp}^{-1} \gtrsim 3{,}000$: A secular growth of $\mathcal{E}_{em}$ continues at the expense of $\mathcal{E}_{k}$.
\end{itemize}

Panel~\ref{fig:energy}(b) shows respectively the parallel, perpendicular thermal energies and the drift energy of ions (protons, and alpha particles). $\mathcal{E}_{th\|, p}$ is growing while $\mathcal{E}_{d, p}$ decreases throughout the entire simulation, except during phase~II, where both energies reach a plateau. Conversely, $\mathcal{E}_{th\perp, p}$ remains constant near zero. Starting from phase~I, $\mathcal{E}_{th\perp, \alpha}$ grows and dominates over the parallel component $\mathcal{E}_{th\|, \alpha}$, which stays close to zero. However, during phase~II, the trend reverses: by the end of phase~II, $\mathcal{E}_{th\perp, \alpha} > \mathcal{E}_{th\|, \alpha}$, and then both energies grow steadily during phase~III, and the trend eventually reverses again at the end of this phase. Throughout all three phases $\mathcal{E}_{d, \alpha}$ decreases, changing slope at each transition.

Panel~\ref{fig:energy}(c) shows the pressure anisotropy (equivalent to temperature anisotropy for near-uniform density) for the different species, defined as $A_{s}\doteq p_{\perp, s}/p_{\|,s}$. We observe that the initial thermal anisotropy for ions is equivalent $A_p=A_{\alpha} = 0.7$. For protons, it decreases slightly during phase~I to reach a plateau during phase~II, and then slightly decreases again during phase~III. Also in this case, these variations are comparable or below $2\%$. In contrast, the pressure anisotropy of the alpha particles with respect to its value in the initial condition decreases during phase~I, increases up to saturation during phase~II, and then decreases again during phase~III. The relative overall change is around $13\%$, indicating significant anisotropy variations.

Panel~\ref{fig:energy}(d) shows the ion drift velocities. At initialization, alpha particles drift about ten times faster than protons. For alpha particles, the drift speed decreases smoothly during phase~I, then drops rapidly during phase~II, and finally continues to decrease almost linearly with a smaller slope during phase~III. In contrast, protons gradually reduce their drift speed throughout the entire simulation without any significant changes in slope.

Although the simulation is evolved up to $t/\Omega_{cp}^{-1} = 15{,}600$, the time series are displayed only up to $t/\Omega_{cp}^{-1} = 6{,}000$, since beyond this point the quantities evolve asymptotically without significant variation.

% Result 2: B field's staking plot 

\begin{figure}[ht!]
\centering
\includegraphics[width=\columnwidth]{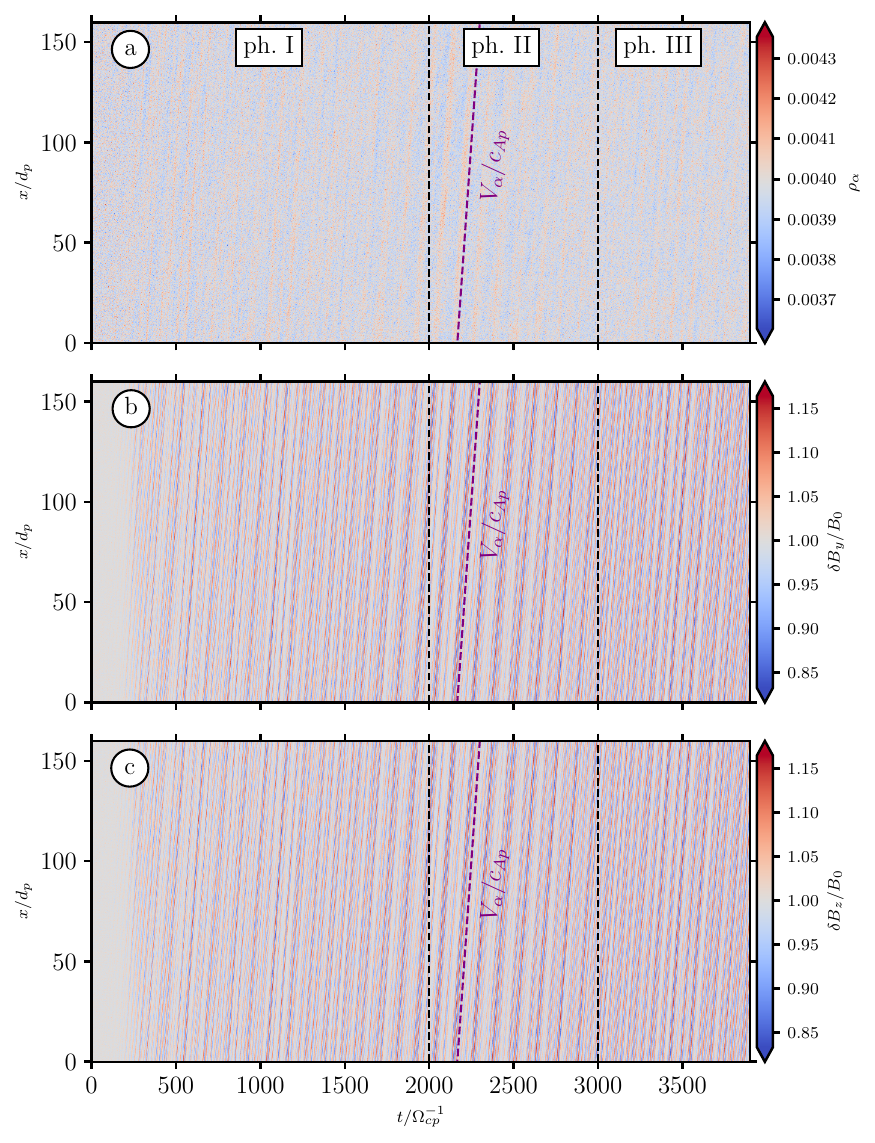}
\caption{Stacked plots on the $x$-$t$ plane of the key quantities of the unstable eigenmode. Panel (a) shows $\rho_{\alpha}$, while panels (b) and (c) show the fluctuations of the transverse magnetic field components, $\delta B_y/B_0$ and $\delta B_z/B_0$, respectively. The purple dashed line indicates the spatio-temporal trajectory of the perturbation, with slope $V_{\alpha}/c_{Ap}$.}
\label{fig:bstack}
\end{figure}

We now visualize the developing unstable modes in the $x$-$t$ plane. Figure~\ref{fig:bstack} presents a series of stacking plots illustrating how different quantities vary in time across our one-dimensional domain. The stacking procedure employs a time cadence of $\tau_{\textrm{Nyquist}}/\omega_{pp}^{-1} = 10^3 \Delta t$, corresponding to a normalized Nyquist frequency of $\omega_{\textrm{Nyquist}}/\Omega_{cp} \approx 8.05$. With $\Omega_{cp}/\omega_{pp} = 100$, this choice ensures that the ion cyclotron frequency is well resolved. The temporal domain is also reduced in this case, truncating part of phase~III to avoid repetitions and to improve readability. Figure~\ref{fig:bstack}(a) shows the density fluctuations of alpha particles, $\rho_{\alpha}$, indicating that the wave has a compressive nature. In panel~\ref{fig:bstack}(b), the fluctuation of the magnetic field in the $x$-direction remains zero throughout the simulation, as the perturbation has no contribution along the direction of propagation. In contrast, the energy is concentrated in the perpendicular directions, where $\delta B_{y}/B_0$ and $\delta B_{z}/B_0$ have non-zero values, confirming that the wave is transverse.
In panel~\ref{fig:bstack}(b) and (c), we plot the spatio-temporal relation of the perturbation as a dashed line whose slope matches $V_{\alpha}/c_{Ap}$. Overall, in panel~\ref{fig:bstack}(c) and (d), it is noticeable that even during phase~I, perturbations appear in $\rho_{\alpha}$ and $\delta B_{y,z}/B_0$ with a similar slope to the main wave, but with higher frequency and shorter temporal separation. In phase~II, the perturbations exhibit lower frequency and larger temporal separation; these are amplified and propagate into the phase~III without significant change.

% Result 3: Power spectral density: omega-k spectrum and FFT vs time

\begin{figure}[ht!]
\centering
\includegraphics[width=\columnwidth]{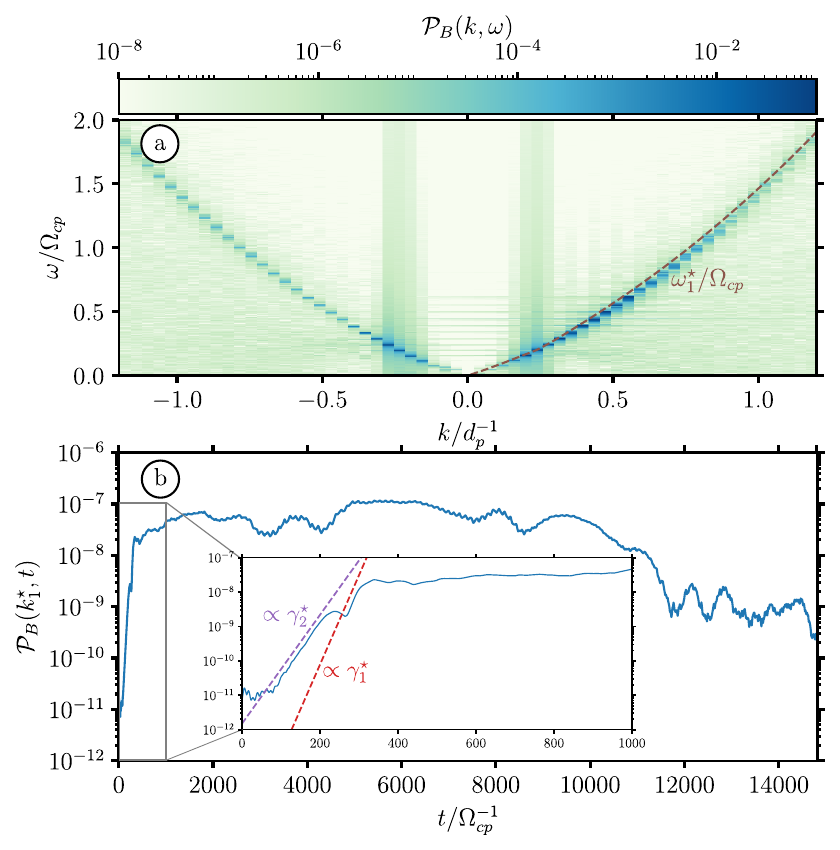}
\caption{Spectral analysis of the numerical simulation's data is compared with that from the LT. Panel~(a) shows the power spectral density in $k$-$\omega$ space (here, $k$ is considered in the parallel direction). The dashed brown line represents the dispersion relation of the fastest growing mode, $\omega_{1}^{\star}$, corresponding to $\theta_{\boldsymbol{k}\boldsymbol{B}}^{\star} = 0$ (see Section~\ref{subsec:linear}). Panel~(b) shows the time evolution of the power spectral density at $k_{1}^{\star}$. In the zoomed-in view, the red dashed line represents the theoretical growth rate of the fastest-growing eigenmode (see Section~\ref{subsec:linear}), whereas the purple dashed line marks an unexpected fluctuation observed in the simulation. See the main text for a detailed discussion of these quantities.}
\label{fig:spectra}
\end{figure}

Wavenumber-frequency spectra are a useful tool to identify the presence of waves and distinguish them from other kind of fluctuations, such as coherent structures \citep[see e.g.,][]{arro2025a, arro2025, papini2021}. The spectral analysis of the kinetic simulation compared with the results from the LT discussed in Section~\ref{subsec:linear} is shown in Figure~\ref{fig:spectra}. Panel~\ref{fig:spectra}(a) shows the measured $k$-$\omega$ spectrum, where we consider $\mathcal{P}_{B}$, the magnetic-field power spectral density (PSD). 
The power is concentrated in the blue region, which matches the dispersion relation derived in Section~\ref{subsec:linear} for the fastest growing mode, $\omega_{1}^{\star}$, propagating parallel to the background magnetic field ($\theta_{\boldsymbol{k}\boldsymbol{B}}^{\star} = 0$).
Panels~\ref{fig:spectra}(b) presents the temporal evolution of the PSD at the wavenumber of the fastest growing eigenmode $k_{\|,1}^{\star}$. The theoretical growth rate, $\gamma_{1}^{\star}$, is shown together with an additional growth rate $\gamma_{2}^{\star}$, which corresponds to the oblique eigenmode. A comprehensive discussion of its interpretation is provided in Section~\ref{subsec:disc-ql}.

% Result 4: VDF alpha

\begin{figure*}[ht!]
\centering
\includegraphics[width=1\textwidth]{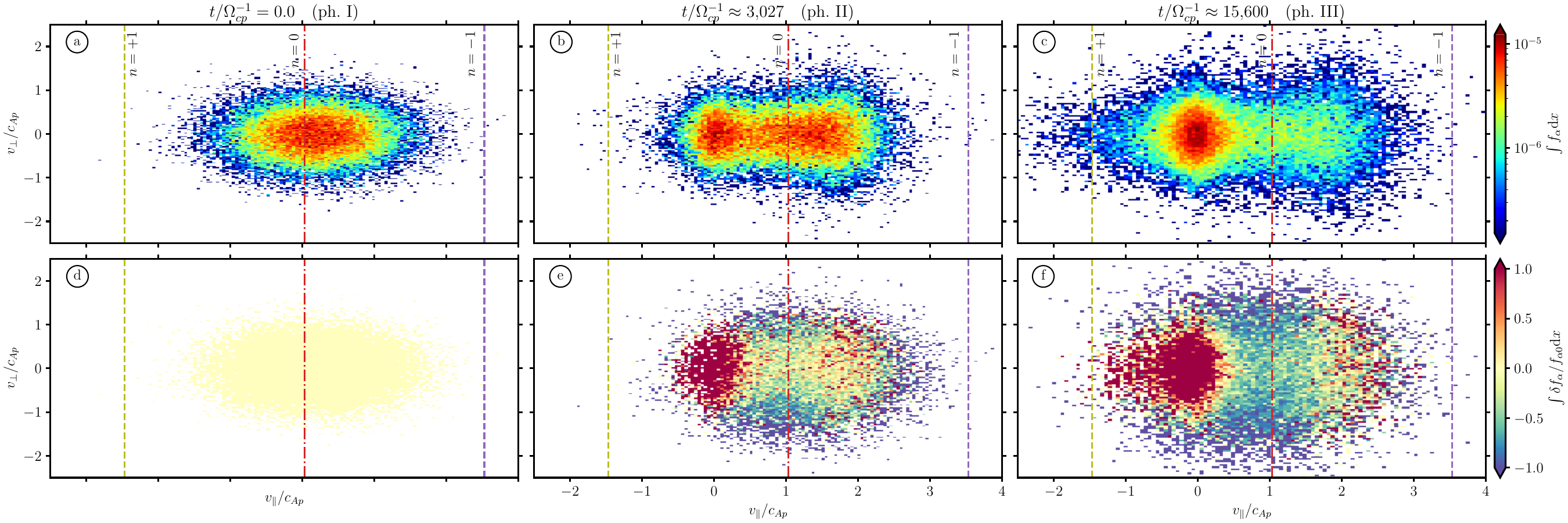}
\caption{Top row: VDFs of the alpha particles throughout the system's evolution. Bottom row: relative variation of the alpha VDFs with respect to the initial $f_{\alpha 0}$ distribution. Across all panels, the resonance lines are indicated as follows: the green dashed line corresponds to the $n = +1$ cyclotron resonance, the red dash-dotted line marks the Landau resonance ($n = 0$), and the purple dashed line indicates the $n = -1$ cyclotron resonance.} 
\label{fig:vdf_alpha_a}
\end{figure*}

In Figure~\ref{fig:vdf_alpha_a}, panels~(a) to (c) show the VDFs of alpha particles at different phases of the system's evolution. Panel~\ref{fig:vdf_alpha_a}(a) shows the alpha VDF at $t/\Omega_{cp}^{-1} = 0.0$, the initialization during the first frame of phase~I (see Section~\ref{sec:setup}).
Panel~\ref{fig:vdf_alpha_a}(b) corresponds to the peak of the phase~II at $t/\Omega_{cp}^{-1} \approx 3{,}027$. In the presence of $n=0$ resonance, the VDF exhibits a thinning along this resonance line, forming two distinct lobes at its extremities. Additionally, the distribution appears to be broader in the parallel direction. Panel~\ref{fig:vdf_alpha_a}(c) displays the VDF during phase~III at $t/\Omega_{cp} \approx 15{,}600$, which marks the end of the simulation. Compared to panel~\ref{fig:vdf_alpha_a}(b), the lobe at $v_{\parallel}/c_{Ap} < 1$ is denser, while the lobe at $v_{\parallel}/c_{Ap} > 1$ is more diffuse and less populated. Furthermore, a newly populated region of phase space emerges near the $n=+1$ resonance line around $v_{\perp}/c_{Ap} \approx 0$.

Panels~\ref{fig:vdf_alpha_a}(d) to (f) show the relative variation of alpha particles VDFs with respect to the initial distribution $f_{\alpha 0} \equiv f_{\textrm{bi-M}, \alpha}$. Here, red regions indicate particle accumulation, blue regions indicate depletion, and yellow regions correspond to stationary areas with no significant change with respect to the initial state. Panel~\ref{fig:vdf_alpha_a}(d) (phase~I) exhibits no discernible variation, as expected, because it is normalized using the same VDF. In panel~\ref{fig:vdf_alpha_a}(e) (phase~II), an accumulation region appears around $v_{\parallel}/c_{Ap} \approx 0$, while depletion regions arise around $v_{\parallel}/c_{Ap} \approx \pm 1$ at $v_{\perp}/c_{Ap} \approx 1$. In panel~\ref{fig:vdf_alpha_a}(f) (phase~III), the accumulation region at $v_{\parallel}/c_{Ap} < 1$ becomes more pronounced, and a new, more diffuse accumulation region emerges at $v_{\parallel}/c_{Ap} > 1$.

\begin{figure}[ht!]
\centering
\includegraphics[width=1\columnwidth]{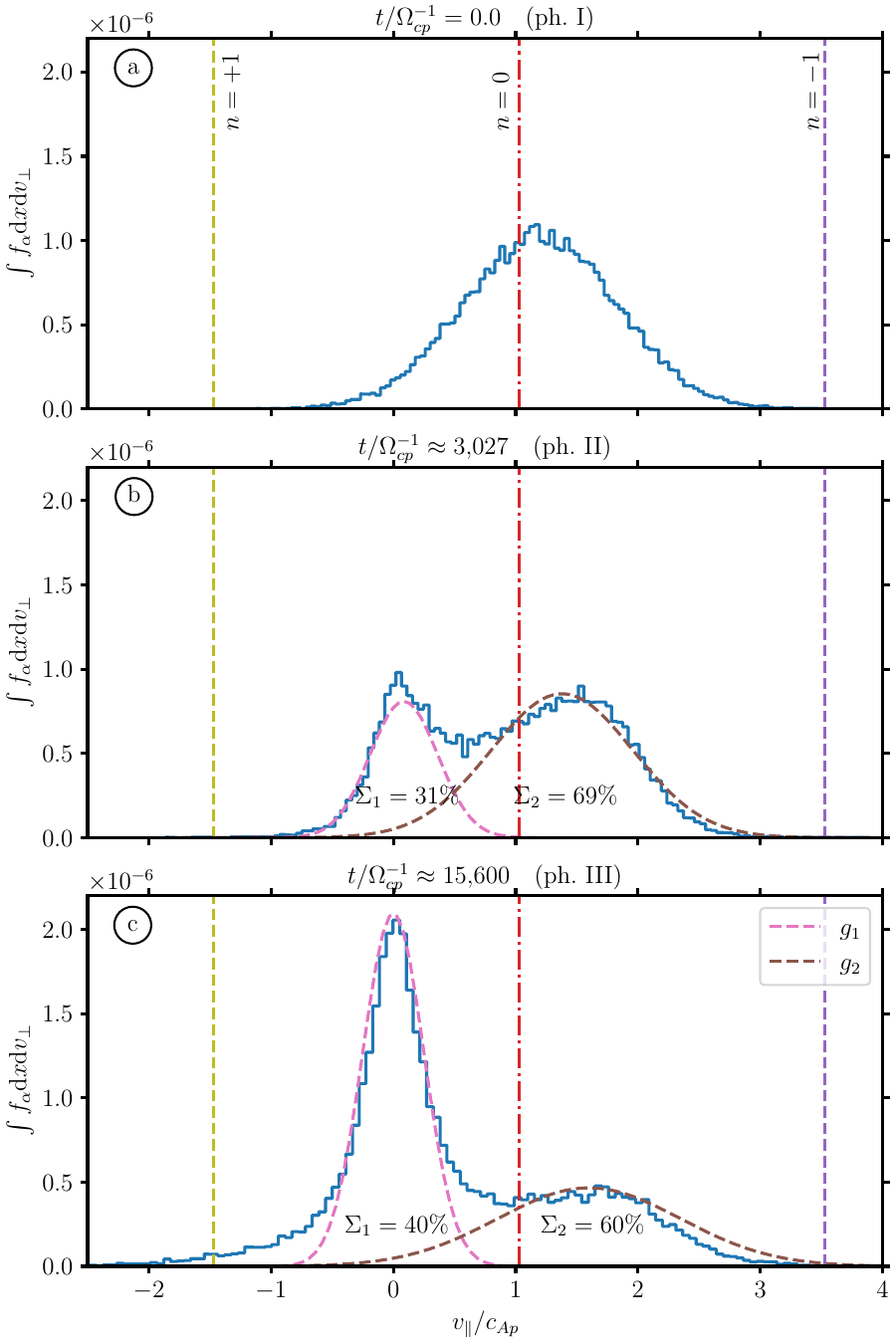}
\caption{VDFs of the alpha particles throughout the system's evolution integrated over $v_{\perp}$. In panels~(b) and (c), the dashed lines represent two Gaussian fits to the distributions. Resonance lines: $n = +1$ (green dashed), $n = 0$ (red dash-dotted), $n = -1$ (purple dashed).} 
\label{fig:vdf_alpha_b}
\end{figure}

Figure~\ref{fig:vdf_alpha_b} shows the alpha VDFs integrated over $v_{\perp}$ to highlight the resonant structures at different phases. Panel~\ref{fig:vdf_alpha_b}(a) presents a Maxwellian distribution centered at $V_{\alpha}/c_{Ap} = 1.2$, consistent with the initial setup. In panel~\ref{fig:vdf_alpha_b}(b), corresponding to phase~II, two distinct peaks appear. These are fitted using Gaussian functions of the form
\begin{equation}
    \begin{aligned}
        g_i(v_{\|}) =\frac{1}{\sqrt{2 \pi \sigma_i^2}} \exp\left(-\frac{(v_{\|}-\mu_i)^2}{2 \sigma_i^2}\right),
    \end{aligned}
    \label{eq:Gauss}
\end{equation}
where the index $i$ denotes the two Gaussian components, $\mu_i$ is the mean (center) of the $i$-th distribution, and $\sigma_i$ is its standard deviation (width). The fitted parameters are $(\mu_1, \sigma_1) = (0.00056, 0.21)$, corresponding to the pink Gaussian, and $(\mu_2, \sigma_2) = (0.83, 1.181)$, corresponding to the brown Gaussian. Because a Gaussian represents a probability density function, the area under each curve corresponds to the fraction of alpha particles in that region and can be used to distinguish between different particle populations. Integrating over $v_{\parallel}$ gives
\begin{equation}
    \begin{aligned}
       \Sigma_{i}= \int_{v_{\|, \textrm{min}}}^{v_{\|, \textrm{max}}} g_i\left(v_{\|}\right) \mathrm{d}v_{\|},
    \end{aligned}
    \label{eq:ndensity}
\end{equation}
yielding $\Sigma_1=31\%$ and $\Sigma_2=69\%$, which represent the fraction of particles contained in each distribution with respect to the total one. We calculate each $\Sigma$ over the interval [$\mu \pm 4\sigma$], which includes more than $99.99\%$ of the distribution, thus providing a precise estimate of the corresponding particle population.

Panel~\ref{fig:vdf_alpha_b}(c) shows the result of applying the same fitting procedure to phase~III. In comparison with phase~II, we observe that the pink distribution becomes sharper, while the brown distribution broadens significantly and extends beyond the resonant lines ($n = \pm 1$), forming high-velocity tails. In this case the fitted Gaussian have parameters $(\mu_1, \sigma_1) = (0.0, 0.25)$, corresponding to the pink Gaussian, and $(\mu_2, \sigma_2) = (1.6, 0.75)$, leading to an area of the pink distribution is $\Sigma_1 = 40\%$ of the total, while the brown distribution area is $\Sigma_2= 60\%$ of the total, reflecting a respective increase and decrease of $9\%$ relative to the phase~II. In this framework, the system consists of an alpha core (pink Gaussian) and a beam (brown Gaussian) with a relative drift of $\Delta v_{\alpha, c-b}/c_{Ap} = 1.8$. The number densities are $n_{\alpha c}/n_{\alpha0} = 40\%$ for the core and $n_{\alpha b}/n_{\alpha0} = 60\%$ for the beam. This configuration is unusual and imbalanced toward the beam, containing more particles than typically expected.

% Result 5: VDF proton
\begin{figure*}[ht!]
\centering
\includegraphics[width=1\textwidth]{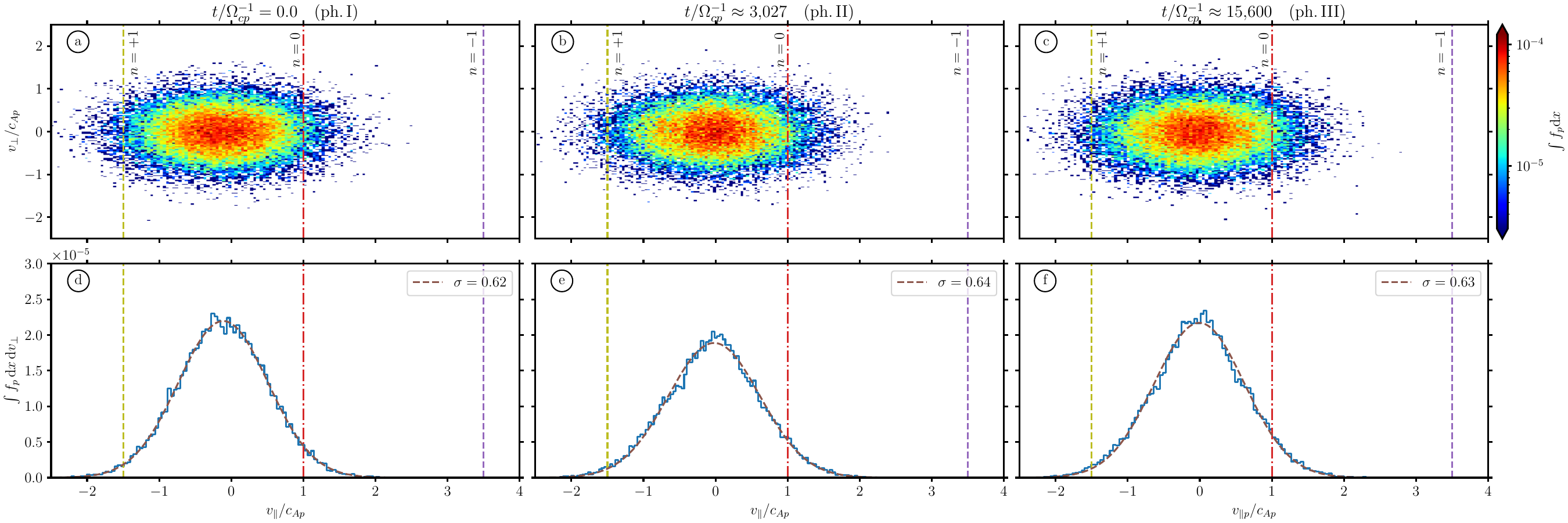}
\caption{Evolution of the proton distribution through the different evolution phases. Top: proton VDF. Bottom: proton distribution function integrated over $v_{\perp}$ plotted vs. $v_{\parallel}$. In panels~(h) and (i), the dashed lines represent a Gaussian fit to the distribution, and $\sigma$ is its variance. Resonance lines: $n = +1$ (green dashed), $n = 0$ (red dash-dotted), $n = -1$ (purple dashed).}
\label{fig:fig9_vdf_proton}
\end{figure*}

In Figure~\ref{fig:fig9_vdf_proton}, we show the evolution of the proton VDF throughout the system's evolution, following the same structure as Figure~\ref{fig:vdf_alpha_a}. Panels~\ref{fig:fig9_vdf_proton}(a) to (c) display the proton VDFs during the different phases, where no substantial changes are observed, and no resonant features develop. In panels~\ref{fig:vdf_alpha_a}(d) to (f), we fit the proton distributions with a single Gaussian profile (brown dashed line) and report their variances. The distribution begins with a standard deviation of $\sigma = 0.62$ in phase~I, slightly broadens to $\sigma = 0.64$ in phase~II, and then narrows again to $\sigma = 0.63$ in phase~III.

% Result 6: transfer rate alpha
\subsection{Field–Particle Correlator (FPC)}\label{subsec:fpc}
\begin{figure*}[ht!]
\centering
\includegraphics[width=1\textwidth]{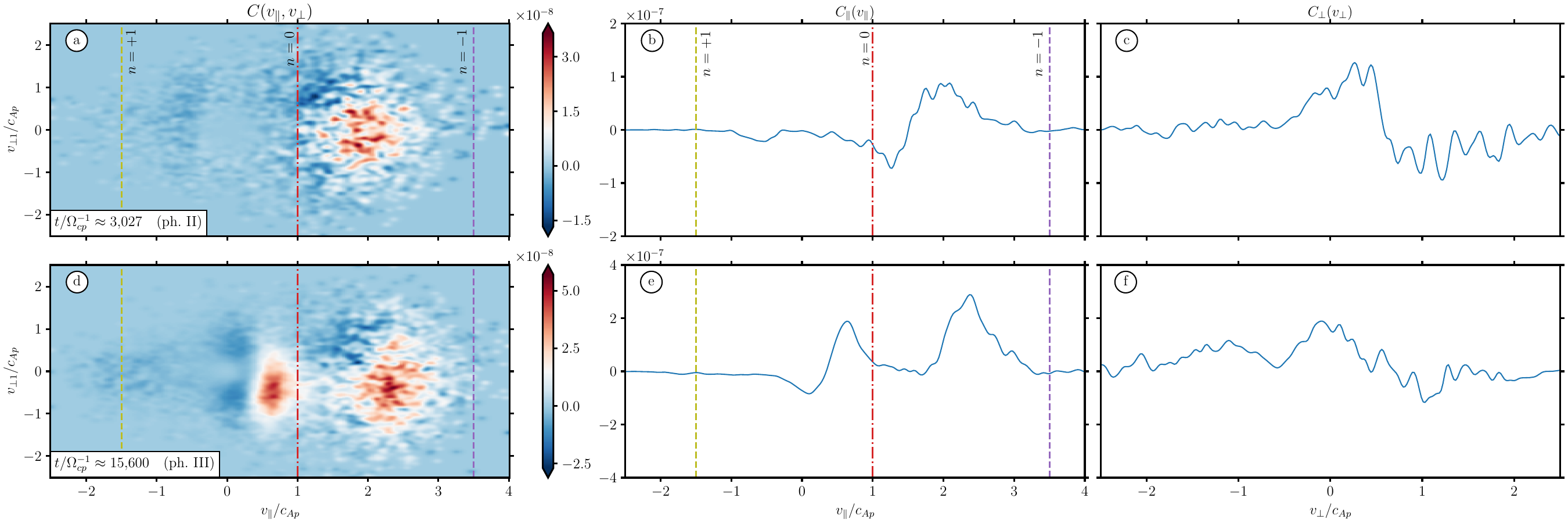}
\caption{Velocity-space signatures of ETR for alpha particles during phase~II (top) and phase~III (bottom). Panel~(a)/(d) shows the ETR, Equation~\eqref{eq:transfer}, in the $v_{\|}$–$v_{\perp}$ space. Panels~(b)/(e) show the ETR as a function of the $v_{\|}$, Equation~\eqref{eq:transferpar}, and panels~(c)/(f) show the ETR as a function of the $v_{\perp}$, Equation~\eqref{eq:transferperp}. Resonance lines: $n = +1$ (green dashed), $n = 0$ (red dash-dotted), $n = -1$ (purple dashed).}
\label{fig:transfer_alpha}
\end{figure*}

After identifying Landau damping in the alpha particle VDF as a consequence of beam-driven instability due to the alpha–proton relative drift, we aim to quantify its contribution to energy transfer between the fields and alpha particles during the phases in which it is active (phase~II and phase~III). Additionally, we investigate whether other, potentially weaker, cyclotron resonances contribute to the energy exchange, even if they are not directly visible in the VDF. To achieve this, we employ the FPC technique introduced by \cite{howesDiagnosingCollisionlessEnergy2017}, which allows us to evaluate the relative contributions of different resonant processes to energy conversion. Following the approach of \cite{jiangVelocityspaceSignaturesResonant2024}, we compute the velocity-space energy density as
\begin{equation}
    \begin{aligned}
       W\left(v_{\|}, v_{\perp},  t\right)=\frac{1}{2} m_{\mathrm{\alpha}} v^2 f_{\alpha}\left(v_{\|}, v_{\perp}, t\right),
    \end{aligned}
    \label{eq:energydensity}
\end{equation}
where $v= (v_{\|}^2+v_{\perp}^2)^{1/2}$ is the magnitude of the particle velocity. The corresponding two-dimensional energy transfer rate (ETR) is given by
\begin{equation}
    \begin{aligned}
       C\left(v_{\|}, v_{\perp}\right)=\frac{\partial W\left(v_{\|}, v_{\perp}, t\right)}{\partial t},
    \end{aligned}
    \label{eq:transfer}
\end{equation}
and is computed using a finite-difference approximation. The temporal increment used in this approximation is $\tau_{\textrm{Nyquist}}/\omega_{pp}^{-1}=10^3\Delta t$, as introduced earlier. By performing partial integration, we obtain the one-dimensional ETR both in the parallel direction,
\begin{equation}
    \begin{aligned}
       C_{\|}\left(v_{\|}\right)=\int_0^{\infty} 2 \pi v_{\perp} C\left(v_{\|}, v_{\perp} \right) \mathrm{d} v_{\perp},
    \end{aligned}
    \label{eq:transferpar}
\end{equation}
and perpendicular direction,
\begin{equation}
    \begin{aligned}       C_{\perp}\left(v_{\perp}\right)=\int_0^{\infty} 2 \pi v_{\|} C\left(v_{\|}, v_{\perp}\right) \mathrm{d} v_{\|}.
    \end{aligned}
    \label{eq:transferperp}
\end{equation}
The integration is carried out employing the Simpson algorithm, owing to its reliability and computational simplicity. The velocity-space energy transfer function $C(v_{\|}, v_{\perp})$, defined in Equation~\eqref{eq:transfer}, is significantly affected by numerical noise. To mitigate this, we convolve this quantity with a Gaussian filter during post-processing to produce a smoother and more interpretable result.

The filtered ETRs are shown in Figure~\ref{fig:transfer_alpha}. Panel~\ref{fig:transfer_alpha}(a) presents $C(v_{\|}, v_{\perp})$ during phase~II. A distinct positive region ($C > 0$) appears at super-Alfv\'enic parallel speeds around $v_{\|}/c_{Ap} \approx 2$, accompanied by a negative region ($C < 0$) near $v_{\|}/c_{Ap} \approx 1$, close to the $n=0$ resonance. This pattern resembles the expected bipolar double-band signature along the $v_{\|}$ direction, as previously described in \citet{jiangVelocityspaceSignaturesResonant2024}. Panel~\ref{fig:transfer_alpha}(b) shows the parallel energy transfer component Equation~\eqref{eq:transferpar}, where the double-band structure is more clearly visible and centered around the Landau resonance ($n=0$). Panel~\ref{fig:transfer_alpha}(c) displays the perpendicular component Equation~\eqref{eq:transferperp}, in this case, the signal has a shape similar to a typical cyclotron-resonance profile \citep{jiangVelocityspaceSignaturesResonant2024}, but it exhibits pronounced oscillations with a larger frequency than expected.

In panel~\ref{fig:transfer_alpha}(d), we show $C(v_{\|}, v_{\perp})$ during phase~III. The double-band signature observed in panel~\ref{fig:transfer_alpha}(a) for $v_{\|}/c_{Ap} > 1$ is still present but appears shifted toward lower parallel velocities, around $v_{\|}/c_{Ap} < 1$. Additionally, a new peak emerges at $v_{\|}/c_{Ap} > 1$, which does not exhibit a sign inversion. The amplitude of this new peak is also noticeably higher compared to the features in panel~\ref{fig:transfer_alpha}(b). In panel~\ref{fig:transfer_alpha}(f), we present $C_{\perp}(v_{\perp})$ for phase~III, which preserves the oscillatory behavior observed earlier and, similarly, does not exhibit any recognizable pattern.

%–––––––––––––––––––––––––––––––––––––––––––––––––––––––––––––––––––%
%                         DISCUSSION SECTION                        %
%–––––––––––––––––––––––––––––––––––––––––––––––––––––––––––––––––––%

\section{Discussion}\label{sec:disc}
In the inner heliosphere, the plasma is inherently turbulent: energy cascades across spatial and temporal scales and ultimately dissipates at kinetic scales, where it contributes to particle heating. In this study, we consider the dissipation at kinetic scale of a low-$\beta$ SW plasma, composed of electrons (modeled by a Maxwellian distribution) and ions, which are divided into protons and alpha particles, both described by drifting bi-Maxwellian distributions. In the electron rest frame, the non-equilibrium protons and alphas drift in opposite directions: alpha particles drift at super-Alfv\'{e}nic speed in the positive direction, while protons drift at one-tenth of the alpha speed in the negative direction.

In this case, SW expansion is neglected, as basic estimates indicate that the timescale of expansion is significantly longer than the non-linear growth time scale of the system (see the asymptotic estimation in Section~\ref{sec:setup}). The setup we employed is similar to that of \citet{tu2002}; therefore, the present work can be regarded as an extension of their analytical study toward a more complete, fully kinetic, non-linear representation.  

\subsection{Implications from LT}\label{subsec:disc-qlt}
We performed a linear analysis of the system using the \textsc{DIS-K} linear solver \citep{lopezGeneralDispersionProperties2021, lopezRalopezhDiskFirst2023}. The system was found to be producing two main branches of unstable eigenmodes: the FM/W quasi-parallel propagating branch ($\boldsymbol{k} \times \boldsymbol{B} = \boldsymbol{0}$), similar to that observed in proton core beam-driven instabilities studied by \citet{pezziniFullyKineticSimulations2024}, and the A/IC obliquely propagating branch ($\boldsymbol{k} \times \boldsymbol{B} \neq \boldsymbol{0}$). The analysis was carried out using both the realistic electron-to-proton mass ratio, $m_p/m_e = 1836$, and a reduced ratio of $m_p/m_e = 100$. We found that varying the mass ratio had a negligible effect on the unstable modes, thereby justifying the use of a reduced mass ratio in numerical simulations to conserve computational resources. Moreover, we studied the damping of the unstable eigenmodes via interactions with protons and alpha particles. Our results suggest that protons can effectively damp the oblique modes through Landau damping, in competition with alpha particles. In contrast, for the quasi-parallel propagating mode, protons are unable to Landau damp the wave, whereas alpha particles can do so over part of the spectrum at lower wavenumbers. At higher wavenumbers, the wave becomes susceptible to cyclotron damping with $n=1$, which reduces the likelihood of cyclotron damping with $n=-1$ by alpha particles.

\subsection{Wave-particles interactions during phase~I and II}\label{subsec:disc-ql}

We employ the PIC code \textsc{ECsim} \citep{lapentaExactlyEnergyConserving2017,bacchiniRelSIMRelativisticSemiimplicit2023} to conduct a fully kinetic simulation in a one-dimensional domain, which is sufficient to fully capture the development of the dominant quasi-parallel instability eigenmode up to the non-linear stage. This geometric choice, however, inherently excludes the study of oblique modes. 

During phase~I, the system remains marginally stable, and noise-induced perturbations begin to grow. In this phase, the energies of protons and alpha particles exhibit similar behavior: the increase in parallel thermal energy is balanced by a decrease in drift energy, while the perpendicular thermal energy remains nearly constant. As time progresses, these fluctuations grow to amplitudes sufficient to cause a clear, linear increase in the system’s global magnetic energy, primarily at the expense of kinetic energy. This marks the onset of phase~II, during which wave growth leads to general plasma cooling and energy redistribution among species. In this phase, protons undergo a relatively quiet transition, with their energy reaching a plateau, while the alpha particles--whose relative drift, with respect to a quite reference frame, remains the main source of free energy--begin to transfer energy from the parallel to the perpendicular direction.
In practice, we are studying a drift-induced instability, where the primary source of free energy arises from the relative drift between ion species. As the system evolves, this relative velocity decreases, indicating that the free energy is being converted into wave energy and redistributed among the different particle species.

We observe that the most unstable eigenmode--numerically identified via LT--becomes active relatively early in the simulation, as shown in panel~\ref{fig:spectra}(b), where linear growth appears for $t/\Omega_{cp}^{-1} \approx 350$. Its effect on the global magnetic and kinetic energy evolution, however, becomes significant only after approximately $10\,\tau_{\mathrm{lin}}$, this delayed response is consistent with our previous studies, e.g., \cite{pezziniFullyKineticSimulations2024}. This suggests that, before the system begins to interact with particles via Landau damping during phase~II, a finite time interval is required to amplify a collection of unstable modes. In addition, the double-slope spectrum visible in panel~\ref{fig:spectra}(b) results from the competition between modes. In particular, $\gamma_{1}^{\star}$ corresponds to the fastest-growing eigenmode that dominates the system’s evolution, as is also evident from the magnetic field fluctuations in panels~\ref{fig:bstack}(b) and (d); whereas $\gamma_{2}^{\star}$ appears to represent the growth rate associated with the projection of the oblique eigenmode onto the parallel direction (see Section~\ref{fig:ktheta}). Finally, the excellent agreement between the simulated spectra and those computed with LT, shown in panel~\ref{fig:spectra}(a), leads us to conclude that the numerically observed eigenmode is the parallel-propagating FM/W eigenmode with RHCP (see Sections~\ref{subsec:pic} and \ref{subsec:disc-qlt}). A comprehensive description of the system requires a two-dimensional geometry, which will be explored in future work.

From the analysis of particle VDFs, we find that alpha particles exhibit a fragmentation of the original distribution around the Landau resonance. This leads to the formation of two distinct sub-populations: the alpha-core and alpha-beam, resulting in a global increase of temperature anisotropy of $13\%$. The alpha-core appears to be roughly Maxwellian and slowly drifting, attaining drift speed $\sim0.2c_{Ap}$, while the alpha-beam is anisotropic in the parallel direction and remains partially connected to the alpha-core, drifting roughly at $1.2c_{Ap}$. At this stage, the alpha-core contains approximately $31\%$ of the total alpha-particles, while the beam accounts for the remaining $69\%$. The process of core-beam production remains incomplete as the core would typically be expected to contain the majority of the particles \citep{marsch1982}; therefore, we expect this process to progress further during phase~III. In contrast, protons do not appear to resonate with the waves. Instead, they accumulate energy in the parallel direction, resulting in a more spread-out distribution.
 
From the ETR analysis, we observe a bipolar double-band signature in the parallel energy transfer rates $C_\|$ (obtained by calculating the field-particle correlator), characteristic of Landau damping. The numerical measurements of $C_{\perp}$ are dominated by low-frequency oscillations that overwhelm the signal. However, panel~\ref{fig:energy}(c) shows that during phase~II the anisotropy clearly increases, suggesting some form of heating mechanism in the perpendicular direction, which could be mediated by cyclotron-resonance of secondary importance.

\subsection{Wave-particles interactions during phase~III}\label{subsec:disc-nl}
After the peak of the growth phase--where both magnetic and kinetic energies reach their extrema--the system reaches a secular growth phase. Beyond this, the trend slightly reverses: magnetic energy is progressively converted back into kinetic energy, leading to particle heating.

Protons continue to convert their drift energy into parallel thermal energy, while their perpendicular thermal energy remains nearly constant, resulting in a progressive decrease in thermal anisotropy. Their distribution function follows a Gaussian shape, with parameters similar to those observed during phase~II.

Alpha particles convert their drift energy into thermal energy; however, the parallel thermal component dominates over the perpendicular one, also resulting in a decrease in thermal anisotropy. 
The Maxwellian alpha-core becomes denser over time, while the alpha-beam appears more diffuse and less dense compared to phase~II.  Moreover, particles seem to be more spread out in phase space relative to phase~II, occupying regions in phase space near the cyclotron resonances $n=\pm 1$. For $n=-1$ the distribution is more collimated around $v_{\perp}\approx0$, whereas for $n=+1$ it appears less collimated, suggesting the presence of some perpendicular heating.

During this phase, the fragmentation of the alpha-particle VDF progressively slows, and for $t \approx 50\tau_{\text{lin}}$ the system reaches a metastable equilibrium, with the alpha-core containing about $40\%$ of the total alpha population and the alpha-beam the remaining $60\%$. This result is similar to the findings of \citet{bruno2024}, which show an equipartition of density, $50\%$ in the alpha-core and $50\%$ in the alpha-beam. The alpha-core is stationary, while the alpha-beam drifts at about $1.8c_{Ap}$, which seems in agreement with \citet{bruno2024}.

ETR analysis shows that, during this phase, the Landau double band signature in $C_{\|}$ shifts at smaller $v_{\|}$, around the $n=0$ resonance, and an additional peak emerges. This behavior likely results from non-linear effects associated with Landau damping, as the $C_{\perp}$ plot does not show any clear signature of cyclotron resonance. Furthermore, the signal in $C_{\perp}$ is substantially noisy, making the identification of resonant features more difficult. 

\section{Conclusion}\label{sec:concl}

In this work, we propose a new mechanism for the formation of an alpha-beam, which arises through the fragmentation of an initially anisotropic alpha-particle distribution via non-linear Landau damping.

We performed a one-dimensional, fully kinetic study of an unstable SW plasma composed of electrons, protons, and alpha particles. The alpha particles are initialized with a drift slightly above the Alfv\'{e}n speed, while ensuring charge and current neutrality. The system is unstable: the relative drift between alpha particles and protons, together with the thermal anisotropy of the ion species, drives the instability. As the instability develops, the ion drift speeds decrease, and the alpha particles undergo perpendicular heating. The resulting electromagnetic fluctuations correspond to parallel-propagating FM/W modes, which are RHCP and resonate with the alpha population through non-linear Landau damping. This process transfers energy from the parallel to the perpendicular direction of the alpha particles, and mostly leading to a fragmentation of the alpha VDF into a core-beam system. Evolving the system for remarkably long time $t \approx 50 \tau_{\text{lin}}$ it reaches a metastable equilibrium where the resulting alpha-beam contains about $60\%$ of the alpha population, while the alpha-core contains the remaining $40\%$. The newly formed alpha-beam exhibits a super-Alfv\'{e}nic drift speed.

The setup of our simulation is inspired by \citet{tu2002}, who proposed a theoretical mechanism for proton-beam generation in a system where protons are organized into three relatively drifting subpopulations. In contrast, our setup includes only one anisotropic proton population, which better reflects typical SW conditions. This work aims to extend \citet{tu2002} theoretical framework and provides new insights into the behavior of multi-species particle systems in the SW.

Furthermore, SolO observations of ion populations \citep{bruno2024} reveal an apparent equipartition of density between alpha-core and alpha-beam components. Interestingly, this partition resembles our simulated configuration with $60\%$-$40\%$ densities.

The main limitation of this study lies in the geometry of the system. Our one-dimensional setup captures the parallel evolution of the unstable eigenmode but neglects oblique ones. To fully characterize the system, future work should extend the simulations to multiple dimensions, thereby enabling the study of oblique instabilities. In addition, we leave for future work a systematic parameter scan of the relative drift speed between protons and alpha particles, along with corresponding observational comparisons, to explore whether different speeds lead to different configurations of the alpha-core plus beam system.

%–––––––––––––––––––––––––––––––––––––––––––––––––––––––––––––––––––%
%                       AKNOLEDGEMENTS SECTION                      %
%–––––––––––––––––––––––––––––––––––––––––––––––––––––––––––––––––––%

\newpage
\section*{Acknowledgements}\label{sec:acknow}
LP\ acknowledges support from a PhD fellowship in fundamental research awarded by the Research Foundation Flanders (FWO), under grant number 11PCB24N. Computational resources and services were provided by the Vlaams Supercomputer Centrum (VSC), funded through Project 2025-15 by the Research Foundation Flanders (FWO) and the Flemish Government--Department of Economy, Science and Innovation (EWI). All data analyses presented in this work were performed using \textsc{Python} programming language \citep{vanrossum1995}, with core libraries including \textsc{NumPy} \citep{harrisArrayProgrammingNumPy2020} for general numerical computations, \textsc{Matplotlib} \citep{hunterMatplotlib2DGraphics2007} for data visualisation, and \textsc{SciPy} \citep{virtanenSciPy10Fundamental2020} for filtering and curve fitting. LP\ gratefully acknowledges Maria Elena Innocenti for insightful discussions on kinetic instabilities.

FB\ acknowledges support from the FED-tWIN programme (profile Prf-2020-004, project ``ENERGY") funded by BELSPO, as well as from the FWO Junior Research Project G020224N granted by the Research Foundation--Flanders (FWO).

ANZ\ gratefully acknowledges financial support from the Belgian Federal Science Policy Office (BELSPO) in the framework of the PRODEX Programme of the European Space Agency (ESA), under contract number 4000147286.

R.A.L. acknowledges the support of ANID Chile through FONDECyT grant No. 1251712.
%–––––––––––––––––––––––––––––––––––––––––––––––––––––––––––––––––––%
%                        AUTHOR CONNTRIBUTIONS                      %
%–––––––––––––––––––––––––––––––––––––––––––––––––––––––––––––––––––%

\section*{Author Contributions}\label{sec:contrib}
LP\ contributed to: Conceptualization; Investigation;
Methodology; Visualization; Writing--original draft; Writing--review \& editing.\\ FB\ contributed to: Mentoring; Software (\textsc{ECsim--RELsim}); Writing--review \& editing.\\ ANZ\ contributed to: Mentoring; Writing--review \& editing.\\ GA\ contributed to: Writing--review \& editing.\\ RAL\ contributed to: Software (\textsc{DIS-K}); Writing--review \& editing.

%–––––––––––––––––––––––––––––––––––––––––––––––––––––––––––––––––––%
%                          DATA AVAILABILITY                        %
%–––––––––––––––––––––––––––––––––––––––––––––––––––––––––––––––––––%

\section*{Data Availability}\label{sec:data}
The data underlying this article are available from the corresponding author upon reasonable request.

\bibliography{References}{}
\bibliographystyle{aasjournal}
\end{document}